
\magnification=\magstephalf
\baselineskip=13.5pt

\vsize=9.5truein
\hsize=6.5truein

\hfuzz=2pt\vfuzz=4pt
\pretolerance=5000
\tolerance=5000
\parskip=0pt plus 1pt
\parindent=16pt
\font\fourteenrm=cmr10 scaled \magstep2
\font\fourteeni=cmmi10 scaled \magstep2
\font\fourteenbf=cmbx10 scaled \magstep2
\font\fourteenit=cmti10 scaled \magstep2
\font\fourteensy=cmsy10 scaled \magstep2
\font\large=cmbx10 scaled \magstep1
\font\vlarge=cmbx10 scaled \magstep3
\font\sans=cmssbx10
\def\bss#1{\hbox{\sans #1}}
\font\bdi=cmmib10

\def\bi#1{\hbox{\bdi #1\/}}

\font\eightrm=cmr8
\font\eighti=cmmi8
\font\eightbf=cmbx8
\font\eightit=cmti8

\font\eightsy=cmsy8
\font\sixrm=cmr6
\font\sixi=cmmi6
\font\sixsy=cmsy6

\def\tenpoint{\def\rm{\fam0\tenrm}%
  \textfont0=\tenrm \scriptfont0=\sevenrm
		      \scriptscriptfont0=\fiverm
  \textfont1=\teni  \scriptfont1=\seveni
		      \scriptscriptfont1=\fivei
  \textfont2=\tensy \scriptfont2=\sevensy
		      \scriptscriptfont2=\fivesy
  \textfont3=\tenex   \scriptfont3=\tenex
		      \scriptscriptfont3=\tenex
  \textfont\itfam=\tenit  \def\it{\fam\itfam\tenit}%
  \textfont\slfam=\tensl  \def\sl{\fam\slfam\tensl}%
  \textfont\bffam=\tenbf  \scriptfont\bffam=\sevenbf
			    \scriptscriptfont\bffam=\fivebf
			    \def\bf{\fam\bffam\tenbf}%
  \normalbaselineskip=20 truept
  \setbox\strutbox=\hbox{\vrule height14pt depth6pt width0pt}%
  \let\sc=\eightrm \normalbaselines\rm}
\def\eightpoint{\def\rm{\fam0\eightrm}%
  \textfont0=\eightrm \scriptfont0=\sixrm
		      \scriptscriptfont0=\fiverm
  \textfont1=\eighti  \scriptfont1=\sixi
		      \scriptscriptfont1=\fivei
  \textfont2=\eightsy \scriptfont2=\sixsy
		      \scriptscriptfont2=\fivesy
  \textfont3=\tenex   \scriptfont3=\tenex
		      \scriptscriptfont3=\tenex
  \textfont\itfam=\eightit  \def\it{\fam\itfam\eightit}%
  \textfont\bffam=\eightbf  \def\bf{\fam\bffam\eightbf}%
  \normalbaselineskip=16 truept
  \setbox\strutbox=\hbox{\vrule height11pt depth5pt width0pt}}
\def\fourteenpoint{\def\rm{\fam0\fourteenrm}%
  \textfont0=\fourteenrm \scriptfont0=\tenrm
		      \scriptscriptfont0=\eightrm
  \textfont1=\fourteeni  \scriptfont1=\teni
		      \scriptscriptfont1=\eighti
  \textfont2=\fourteensy \scriptfont2=\tensy
		      \scriptscriptfont2=\eightsy
  \textfont3=\tenex   \scriptfont3=\tenex
		      \scriptscriptfont3=\tenex
  \textfont\itfam=\fourteenit  \def\it{\fam\itfam\fourteenit}%
  \textfont\bffam=\fourteenbf  \scriptfont\bffam=\tenbf
			     \scriptscriptfont\bffam=\eightbf
			     \def\bf{\fam\bffam\fourteenbf}%
  \normalbaselineskip=24 truept
  \setbox\strutbox=\hbox{\vrule height17pt depth7pt width0pt}%
  \let\sc=\tenrm \normalbaselines\rm}

\def\today{\number\day\ \ifcase\month\or
  January\or February\or March\or April\or May\or June\or
  July\or August\or September\or October\or November\or December\fi
  \space \number\year}
\newcount\secno      
\newcount\subno      
\newcount\subsubno   
\newcount\appno      
\newcount\tableno    
\newcount\figureno   
\normalbaselineskip=15 truept
\baselineskip=15 truept
\def\title#1#2
   {\vglue 1cm
   {\baselineskip=24 truept
    \pretolerance=10000
    \raggedright
    \centerline{{\vlarge #1}}\smallskip\centerline{{\vlarge #2}}}}
\def\author#1
  {\pretolerance=10000
    \centerline{ {\bsc By #1}}}
\def\address#1
   {\smallskip
    \centerline{ \sl #1}}
\def\shorttitle#1
   {\vfill
    \noindent \rm Short title: {\sl #1}\par
    \medskip}
\def\pacs#1
   {\noindent \rm PACS number(s): #1\par
    \medskip}
\def\jnl#1
   {\noindent \rm Submitted to: {\sl #1}\par
    \medskip}
\def\date
   {\noindent Date: \today\par
    \medskip}
\def\beginabstract
   {\vskip 1cm\baselineskip=12pt
    \noindent \rm}
\def\keyword#1
   {\bigskip
    \noindent {\bf Keyword abstract: }\rm#1}
\def\endabstract
{\centerline{\vbox{\hrule width12cm }}}

\def\entry#1#2#3
   {\noindent
    \hangindent=20pt
    \hangafter=1
    \hbox to20pt{#1 \hss}#2\hfill #3\par}
\def\subentry#1#2#3
   {\noindent
    \hangindent=40pt
    \hangafter=1
    \hskip20pt\hbox to20pt{#1 \hss}#2\hfill #3\par}
\def\section#1
   {
    \vskip0pt plus-.1\vsize\vskip24pt plus12pt minus6pt
    \subno=0 \subsubno=0 \exno=0 \eqnno=0
    \global\advance\secno by 1
    \centerline{ {\bf \the\secno. #1}}
    \nobreak\bigskip\noindent}
\def\subsection#1
   {
    \vskip12pt plus6pt minus3pt
    \bigbreak
    \global\advance\subno by 1
    \subsubno=0
    \centerline{\sl \the\secno.\the\subno. #1}
    \nobreak
    \medskip}
\def\subsubsection#1
   {\vskip-\lastskip
    \vskip20pt plus6pt minus6pt
    \bigbreak
    \global\advance\subsubno by 1
    \centerline{\sl \the\secno.\the\subno.\the\subsubno. #1\null. }}
\def\appendix#1
   {\vskip0pt plus.1\vsize\penalty-250
    \vskip0pt plus-.1\vsize\vskip24pt plus12pt minus6pt
    \subno=0
    \global\advance\appno by 1
    \noindent {\large Appendix \the\appno\qquad #1\par}
    \bigskip
    \noindent}
\def\subappendix#1
   {\vskip-\lastskip
    \vskip36pt plus12pt minus12pt
    \bigbreak
    \global\advance\subno by 1
    \noindent {\sl \the\appno.\the\subno. #1\par}
    \nobreak
    \medskip
    \noindent}

\def\tabcaption#1
   {\global\advance\tableno by 1
    \noindent {\bf Table \the\tableno.} \rm#1\par
    \bigskip}
\def\figures
   {\vskip 1truein
    \centerline{\large Figure captions}
    \bigskip}
\def\figcaption#1
   {\global\advance\figureno by 1
    \noindent {\bf Figure \the\figureno.} \rm#1\par
    \bigskip}
\def\references
     {\vskip .5truein 
     {\centerline{ \large References}}
      \parindent=0pt
      \bigskip}
\def\refjl#1#2#3#4
   {\hangindent=16pt
    \hangafter=1
    \rm #1
   {\frenchspacing\sl #2
    \bf #3}
    #4\par}
\def\refbk#1#2#3
   {\hangindent=16pt
    \hangafter=1
    \rm #1
   {\frenchspacing\sl #2}
    #3\par}
\def\numrefjl#1#2#3#4#5
   {\parindent=40pt
    \hang
    \noindent
    \rm {\hbox to 30truept{\hss #1\quad}}#2
   {\frenchspacing\sl #3\/
    \bf #4}
    #5\par\parindent=16pt}
\def\numrefbk#1#2#3#4
   {\parindent=40pt
    \hang
    \noindent
    \rm {\hbox to 30truept{\hss #1\quad}}#2
   {\frenchspacing\sl #3\/}
    #4\par\parindent=16pt}

\def\frac#1#2{{#1 \over #2}}

\def\d{{\rm d}}
\def\e{{\rm e}}
\def\i{\ifmmode{\rm i}\else\char"10\fi}
\def\case#1#2{{\textstyle{#1\over #2}}}
\def\boldrule#1{\vbox{\hrule height1pt width#1}}

\def\etal{{\it et al\/}.}
\catcode`\@=11
\def\ind{\hbox to 5pc{}}
\def\eq(#1){\hfill\llap{(#1)}}

\def\deqn#1{\displ@y\halign{\hbox to \displaywidth
    {$\@lign\displaystyle##\hfil$}\crcr #1\crcr}}
\def\indeqn#1{\displ@y\halign{\hbox to \displaywidth
    {$\ind\@lign\displaystyle##\hfil$}\crcr #1\crcr}}
\def\indalign#1{\displ@y \tabskip=0pt
  \halign to\displaywidth{\ind$\@lign\displaystyle{##}$\tabskip=0pt
    &$\@lign\displaystyle{{}##}$\hfill\tabskip=\centering
    &\llap{$\@lign##$}\tabskip=0pt\crcr
    #1\crcr}}
\catcode`\@=12



\def\PRL{Phys. Rev. Lett.}

\def\bfsigma{\bi{\char'33}}

\def\bfpsi{\bi{\char'40}}


\font\sc=cmcsc10
\font\bsc=cmcsc10 scaled \magstep1

\def\p{Pain\-lev\'e}\def\bk{B\"ack\-lund}
\def\d{{\rm d}}\def\e{{\rm e}}\def\i{{\rm i}}

\font\bdi=cmmib10
\def\bi#1{\hbox{\bdi #1\/}}

\def\tfr#1#2{{\tx{#1\over#2}}}

\newcount\refno
\def\ref#1#2#3#4#5{\vskip.9pt\global\advance\refno by 1
\item{[{\bf\the\refno}]\ }{\rm#1}, {\it#2}, {\bf#3} (#4) #5}

\def\sam{Stud.\ Appl.\ Math.}

\def\jpa{J.\ Phys.\ A: Math.\ Gen.}
\def\jmp{J.\ Math.\ Phys.}

\def\~#1{{\bf\tilde{\mit#1}}}


\def\secn{\the\secno}

\font\sit=cmti9


\nopagenumbers
\def\sch{Schr\"odinger}

\def\and{\qquad {\rm and}\qquad}

\def\tfr#1#2{{\tx{#1\over#2}}}

\def\pde{partial differential equa\-tion}
\def\pdes{partial differential equa\-tions}
\def\ode{ordinary differential equa\-tion}

\def\eq{equa\-tion}

\def\d{{\rm d}}\def\e{{\rm e}}\def\i{{\rm i}}

\def\sech{\mathop{\rm sech}\nolimits}

\font\teneuf=eufm10 \font\seveneuf=eufm7 \font\fiveeuf=eufm5
\newfam\euffam
\textfont\euffam=\teneuf \scriptfont\euffam=\seveneuf
   \scriptscriptfont\euffam=\fiveeuf

\newcount\exno
\newcount\secno   
\newcount\subno   
\newcount\subsubno   
\newcount\figno
\newcount\tableno

\def\boldrule#1{\vbox{\hrule height1pt width#1}}

\def\bline#1{\boldrule{#1truein}}

\def\Table#1#2#3{\vskip-\lastskip
    \vskip4pt plus2pt minus2pt
    \bigbreak \global\advance\tableno by 1
		 \vbox{\centerline{\bf Table \the\secno.\the\subno%
		 \uppercase\expandafter{\romannumeral\the\tableno}}\smallskip
		      \centerline{\sl #1}
 {$$\vbox{\offinterlineskip\tabskip=0pt
       \bline{#2}
       \halign to#2truein{#3}
       \bline{#2}}
   $$}}}

\newbox\strutbox
\setbox\strutbox=\hbox{\vrule height10pt depth 4.5pt width0pt}

\def\defn#1{\vskip-\lastskip
    \vskip4pt plus2pt minus2pt
    \bigbreak
    \global\advance\exno by 1
    \noindent {{\bf Definition\ \the\secno.\the\subno.\the\exno}\enskip
{\rm#1\/}}\smallskip}
\def\example#1{\vskip-\lastskip
    \vskip4pt plus2pt minus2pt \caseno=0
    \bigbreak \global\advance\exno by 1
    \noindent {{\bf Example\ \the\secno.\the\subno.\the\exno}
\enskip{\rm#1}}\smallskip}
\def\exercise#1{\vskip-\lastskip
    \vskip4pt plus2pt minus2pt \caseno=0
    \bigbreak \global\advance\exno by 1
    \noindent {{\bf Exercise\ \the\secno.\the\subno.\the\exno}
\enskip{\rm#1}}\smallskip}
\def\thm#1{\vskip-\lastskip
    \vskip4pt plus2pt minus2pt \caseno=0
    \bigbreak \global\advance\exno by 1
    \noindent {{\bf Theorem\ \the\secno.\the\subno.\the\exno}\enskip
    {\sl#1\/}}\smallskip}
\def\lem#1{\vskip-\lastskip
    \vskip4pt plus2pt minus2pt \caseno=0
    \bigbreak \global\advance\exno by 1
    \noindent {{\bf Lemma\ \the\secno.\the\subno.\the\exno}\enskip
    {\it#1\/}}\smallskip}
\def\remark{\vskip-\lastskip
    \vskip4pt plus2pt minus2pt \caseno=0
    \bigbreak \global\advance\exno by 1
    \noindent {{\bf Remark\ \the\secno.\the\subno.\the\exno}\enskip}}
\def\remarks{\vskip-\lastskip
    \vskip4pt plus2pt minus2pt \caseno=0
    \bigbreak \global\advance\exno by 1
    \noindent {{\bf Remarks\ \the\secno.\the\subno.\the\exno}\enskip}}

\def\=#1{{\bf\bar{\mit#1}}}
\def\^#1{{\bf\hat{\mit#1}}}
\def\~#1{{\bf\tilde{\mit#1}}}

\def\cc#1{\kappa_{#1}}

\def\hbb#1#2#3{\qquad{\hbox to 100pt{$#1$\hfill}}{\hbox to
200pt{$#2$\hfill}}\hfill#3}
\def\ra{{\hbox to 100pt{\hfill}}\Rightarrow\qquad}

\newcount\remno
\def\tfr#1#2{{\textstyle{#1\over#2}}}
\def\rem#1{\global\advance\remno by 1
\item{\the\remno.\enskip}#1}

\newcount\eqnno
\def\sen{\the\secno}
\def\secsub{\the\secno.\the\subno}
\def\eqn#1{\global\advance\eqnno by 1
	   \eqno(\sen.\the\eqnno)
	   \expandafter \xdef\csname #1\endcsname
	   {\sen.\the\eqnno}\relax }
\def\eqnn#1{\global\advance\eqnno by 1
	   (\sen.\the\eqnno)
	   \expandafter \xdef\csname #1\endcsname
	   {\sen.\the\eqnno}\relax }
\def\eqnm#1#2{\global\advance\eqnno by 1
	   (\sen.\the\eqnno{\it#2})
	   \expandafter \xdef\csname #1\endcsname
	   {\sen.\the\eqnno}\relax }
\def\eqnr#1{(\sen.\the\eqnno{\it#1})}

\newcount\caseno
\def\case#1{\vskip-\lastskip
    \vskip4pt plus2pt minus2pt
    \bigbreak \global\advance\caseno by 1
\noindent\underbar{{\sc Case}
\the\secno.\the\subno.\the\caseno}\enskip{#1}. }

\def\Case#1{\vskip-\lastskip
    \vskip4pt plus2pt minus2pt
    \bigbreak \global\advance\caseno by 1
\noindent {\sc Case}\
\the\secno.\the\subno.\the\exno{\romannumeral\the\caseno}\enskip
\underbar{#1}.\quad}

\def\hide#1{}

\def\cite#1{[#1]}
\def\PSI{\bfpsi}
\def\DELTA{{\bf\Delta}}

\def\pa{\partial}
\def\tz{\~z}
\def\tzeta{\~\zeta}
\def\pii{P$_{II}$}
\def\piii{P$_{III}$}

\def\pv{P$_{V}$}
\def\bt{B\"ack\-lund trans\-forma\-tion}
\def\bts{B\"ack\-lund trans\-forma\-tions}

\def\Fz{F_{\zeta}}
\def\Fzz{F_{\zeta\zeta}}
\def\Fzzz{F_{\zeta\zeta\zeta}}
\def\ex#1{\exp\left(#1\right)}

\headline={\ifnum\pageno>1
\ifodd\pageno\rightheadline\else\leftheadline\fi
\else\hfil\fi}
\def\shortdate{}
\def\rightheadline{\tenrm{\eightrm\shortdate} \hfil{\sit On a
2+1-dimensional
sine-Gordon system}\hfil\folio}
\def\leftheadline{\tenrm{\eightrm\shortdate} \hfil {\sit
P.A.\ Clarkson, E.L.\
Mansfield \& A.E.\ Milne}\hfil\folio}

\title{Symmetries and Exact Solutions of}{a 2+1-dimensional Sine-Gordon
System}
\vskip 1truecm
\author{Peter A.\ Clarkson, Elizabeth L.\ Mansfield and Alice
E.\ Milne}
\address{Department of Mathematics, University of Exeter,
 Exeter, EX4 4QE, U.K.}
\beginabstract
We investigate the classical and nonclassical reductions of the
$2+1$-dimensional
sine-Gordon system of Konopelchenko and Rogers, which is a strong
generalisation of
the sine-Gordon equation. A family of solutions obtained as a
nonclassical reduction
involves a decoupled sum of solutions of a generalised, real, pumped
Maxwell-Bloch
system. This implies the existence of families of solutions, all
occurring as a
decoupled sum, expressible in terms of the second, third and fifth
Painlev\'e
transcendents, and the sine-Gordon equation. Indeed, hierarchies of
such solutions
are found, and explicit trans\-forma\-tions connecting members of each
hierarchy are
given.

By applying a known B\"acklund trans\-forma\-tion for the system to the
new solutions
found, we obtain further families of exact solutions, including some
which are
expressed as the argument and modulus of sums of products of Bessel
functions with
arbitrary coefficients.

Finally, we prove the sine-Gordon system has the Painlev\'e property,
which
requires the usual test to be modified, and derive a non-isospectral
Lax pair for
the generalised, real, pumped Maxwell-Bloch system.

\smallskip
\endabstract

\section{Introduction}
There is much current interest in $2+1$-dimensional equations, i.e.\
equations with two spatial and one temporal variables, which are
solvable by
inverse scattering and thus are completely integrable (cf., Ablowitz
\&\ Clarkson
1991; Konopelchenko 1993). Notable examples include the
Kadomtsev-Petviashvili
(KP), Davey-Stewartson (DS) and Nizhnik-Veselov-Novikov (NVN)
equations.
The DS and NVN equations are ``strong'' generalisations of the
nonlinear \sch\
equation and Korteweg-de Vries (KdV) equation respectively, since the
spatial
variables arise on an equal basis. In contrast, the KP equation is {\it
not} a
strong generalisation of the KdV equation.  Further, the DS and NVN
equations admit
coherent structure solutions (Boiti \etal\ 1988; Athorne \&\ Nimmo
1991,
respectively); these are exponentially decaying solutions, often
referred to as {\it
dromions} (Fokas \&\ Santini 1989, 1990).

Recently Konopelchenko \&\ Rogers (1991, 1993) have re-derived the
system of
equations
$$\eqalignno{\left({\Theta_{XT} \over \sin\,\Theta}\right)_X
-\left({\Theta_{YT} \over \sin\Theta}\right)_Y
-{\Theta_Y\Phi_{XT}-\Theta_X\Phi_{YT}\over \sin^2\Theta}&=
0,&\eqnm{eqKR}{a}\cr
\left({\Phi_{XT} \over \sin\Theta}\right)_X
-\left({\Phi_{YT} \over \sin\Theta}\right)_Y
-{\Theta_Y\Theta_{XT}-\Theta_X\Theta_{YT}\over \sin^2\Theta}&=0,
&\eqnr{b}\cr}$$
within the context of a novel class of $2+1$-dimensional equations
generalising
work on infinitesimal \bts\ due to Loewner (1952). The system (\eqKR)
is a ``strong" $2+1$-dimensional generalisation of the sine-Gordon
equation
$$\theta_{zt}=\sin\theta,\eqn{eqSG}$$
since reducing it via $\Phi=0$, $\Theta(X,Y,t)=\theta(z,t)$,
where $z=\mu_1 X+\mu_2Y$, with $\mu_1$ and $\mu_2$ constants, not both
zero,
and $t=T$, yields (\eqSG), i.e., the reduction to the sine-Gordon
equation occurs for either of the spatial variables.

The system (\eqKR) is also obtained as a special reduction of an
integrable
$2+1$-dimensional Toda lattice scheme (Rogers 1993). An auto-\bt\ and
some exact
coherent structure solutions for this system are given in Konopelchenko
\etal\
(1992); see also Konopelchenko (1993). Further coherent structure
solutions for
(\eqKR) are given by Schief (1992) and other solitonic solutions are
given by Nimmo
(1992, 1993).

The special case of (\eqKR) when $\Phi=\Theta$ is the $2+1$-dimensional
sine-Gordon equation
$$\left({\Theta_{XT} \over \sin\,\Theta}\right)_X
-\left({\Theta_{YT} \over \sin\Theta}\right)_Y
-{\Theta_Y\Theta_{XT}-\Theta_X\Theta_{YT}\over \sin^2\Theta}=
0,\eqn{eqKRa}$$
which was also written down by Konopelchenko \&\ Rogers (1991).
Remarkably,
the $2+1$-dimensional sine-Gordon equation (\eqKRa) appears in the work
of
Darboux on classical differential geometry, Darboux (1887--1896), as
noted by
Johnson \etal\ (1994).

We remark that (\eqKRa) is {\it not} the ``natural''
$2+1$-dimensional generalisation of the sine-Gordon equation (\eqSG)
given by
$$ \Theta_{TT} - \Theta_{XX}-\Theta_{YY} = \sin\Theta.\eqn{eqSGII}$$
In fact, there is considerable evidence suggesting that (\eqSGII) is
{\it not} completely integrable, i.e.\ solvable by inverse scattering
(cf.,
Clarkson 1986{\it a}).

Making the gauge trans\-forma\-tion
$$\Phi_{xt} = 2v_{xt} \sin\Theta + \Theta_{xt}\cos\Theta,\qquad
\Phi_{yt} = -2v_{yt} \sin\Theta - \Theta_{yt}\cos\Theta,\eqn{eqgauge}$$
with $x=\tfr12(X+Y)$, $y=\tfr12(X-Y)$, $t=T$ and
$u(x,y,t)=\tfr12\Theta(X,Y,T)$
to (\eqKR) yields the following system of equations
(see Konopelchenko \&\ Dubrovsky 1993; Nimmo 1993),
$$\eqalignno{
\Delta_1 &\equiv u_{xyt} + u_x v_{yt} + u_y v_{xt}=
0,&\eqnm{eqgsg}{a}\cr
\Delta_2 &\equiv v_{xy}- u_x u_y=0,&\eqnr{b}\cr}$$
and it is this system which we study. The Lax pair for (\eqgsg) is
given by
$$\eqalignno{
\psi_{1,x} + u_x \psi_2 &= 0,&\eqnm{gsglaxi}{a}\cr
\psi_{2,y} - u_y \psi_1 &= 0,&\eqnr{b}\cr
\psi_{1,yt} + u_y \psi_{2,t} + v_{yt}\psi_1 &= 0,&\eqnm{gsglaxii}{b}\cr
\psi_{2,xt} - u_x \psi_{1,t} + v_{xt}\psi_2 &= 0,&\eqnr{b}\cr}$$
since $\psi_{1,xyt}=\psi_{1,ytx}$ and $\psi_{2,xyt}=\psi_{2,ytx}$
if and only if $u$ and $v$ satisfy (\eqgsg). We remark that the
spectral part of
the Lax pair (\gsglaxi) is the same as that for the DSI equation,
which was solved by inverse scattering by Ablowitz \&\ Haberman (1975)
and Fokas \&\
Ablowitz (1984). By eliminating $v_t$ in (\eqgsg{\it a}) one obtains,
$$ u_{xyt}+m_1(y,t) u_x+m_2(x,t) u_y+u_x\partial^{-1}_x(u_{x}u_y)_t
+u_y\partial^{-1}_y(u_xu_{y})_t=0,\eqn{KDUeqn}$$
where $\partial^{-1}_xf(x)=\int_{-\infty}^xf(x_1)\,\d x_1$,
$\partial^{-1}_yf(y)=\int_{-\infty}^yf(y_1)\,\d y_1$,
and $m_1(y,t)=\lim_{x\to-\infty}v_y(x,y,t)$,
$m_2(x,t)=\lim_{y\to-\infty}v_x(x,y,t)$ are functions of
integration. The inverse scattering transform
and initial value problem for this equation was studied by
Konopelchenko \&\
Dubrovsky (1993; Dubrovsky \&\ Konopelchenko 1993) for the case when
the boundary
values $m_1$ and $m_2$ in (\KDUeqn) are constants.

In \S\S2 and 3 of this paper, we consider reductions of the system
(\eqgsg)
using the classical Lie method and the so-called ``nonclassical
method'' due to
Bluman \&\ Cole (1969), respectively. Using this nonclassical method,
we
obtain a family of solutions involving a decoupled sum of solutions of
a real
generalised Maxwell-Bloch system. This family includes as special cases
solutions
expressible in terms of solutions of the sine-Gordon equation (\eqSG)
and the
second, third and fifth Painlev\'e equations, which are discussed in
\S4. Indeed,
hierarchies of such solutions are given, and the trans\-forma\-tions
connecting
members of each hierarchy are given explicitly. In \S5 we apply a known
\bt\ (5.2) for the system (\eqgsg) to our new exact solutions in
some simple cases, to obtain further families of solutions including
solutions expressed in terms of expansions of exponential or
Bessel functions with arbitrary coefficients.

In \S6, we demonstrate that the sine-Gordon system (\eqgsg) has the
Painlev\'e
property due to Weiss \etal\ (1983); this requires a modification of
the usual
procedure given in Weiss \etal\ (1983). Further we discuss
the integrability of the generalised Maxwell-Bloch system which we
obtained
as a nonclassical reduction. Finally in \S7 we briefly discuss our
results.

\section{Classical Method}
The classical method for finding symmetry reductions of \pdes\ is the
Lie group
method of infin\-ites\-imal trans\-forma\-tions (cf., Olver 1993).
Though this
method is entirely algorithmic, it often involves a large amount of
tedious
algebra and auxiliary calculations which can become virtually
unmanageable if
attempted manually, and so symbolic manipulation programs have been
developed,
for example in MACSYMA, MAPLE, MATHEMATICA, and REDUCE, to facilitate
the
calculations. An excellent survey of the different packages presently
available
and a discussion of their strengths and applications is given by
Hereman
(1994). In this paper we the associated system of determining equations
was
generated using the MACSYMA program {\tt symmgrp.max} (Champagne
\etal\ 1991).
Subsequently this system was analysed interactively using the MAPLE
package {\tt
diffgrob2}, and the Reid strategy for controlling expression swell was
employed
(cf.\ Clarkson \&\ Mansfield 1994{\it c}; Mansfield 1993).

To apply the classical method to the system ({\eqgsg}) we consider the
one-parameter Lie group of infinitesimal trans\-forma\-tions in
$(x,y,t,u,v)$ given by
$$\eqalignno{
\~{x} &={x}+ \varepsilon {\xi_1}(x,y,t,u,v) + O(\varepsilon^2),
&\eqnm{eqinftr}{a}\cr
\~{y} &={y}+ \varepsilon {\xi_2}(x,y,t,u,v) + O(\varepsilon^2),
&\eqnr{b}\cr
\~{t} &= {t} + \varepsilon {\xi_3}(x,y,t,u,v) + O(\varepsilon^2),
&\eqnr{c} \cr
\~{u} &= {u} + \varepsilon {\phi_1}(x,y,t,u,v) + O(\varepsilon^2),
&\eqnr{d}\cr
\~{v} &= {v} + \varepsilon {\phi_2}(x,y,t,u,v) + O(\varepsilon^2),
&\eqnr{e}\cr}$$
where $\varepsilon$ is the group parameter. Then one requires that
this trans\-forma\-tion leaves invariant the set
$${\cal S}_{\DELTA} \equiv \left\{u(x,y,t),v(x,y,t):
\Delta_1(u,v)=0,\Delta_2(u,v)=0\right\},\eqn{Sdelta}$$ of solutions of
the system
(\eqgsg). This yields an overdetermined, linear system of equations for
the
infin\-ites\-imals ${\xi_i}(x,y,t,u,v)$, $i=1,2,3$, and
${\phi_j}(x,y,t,u,v)$,
$j=1,2$.
The associated Lie algebra
of infin\-ites\-imal symmetries is the set of vector fields of the form
$${\bf v} = \xi_1{\partial_x} +\xi_2{\partial_y} + \xi_3{\partial_t} +
\phi_1{\partial_u} + \phi_2{\partial_v}, \eqn{eqvf}$$
where $\partial_x \equiv{\partial/\partial x}$ etc.
Having determined the infinitesimals, the symmetry variables are found
by
solving the invariant surface conditions
$$\eqalignno{\psi_1&\equiv\xi_1u_x +\xi_2u_y+
 \xi_3u_t- \phi_1=0,&\eqnm{insc}{a}\cr
\psi_2&\equiv\xi_1v_x +\xi_2v_y+ \xi_3v_t- \phi_2=0.&\eqnr{b}\cr}$$

Applying the classical method to the system ({\eqgsg}) yields a system
of twenty-one
one-term equations which have the solution
$$\xi_1 =f_1(x),\qquad \xi_2=f_2(y),\qquad \xi_3=f_3(t),\qquad
\phi_1=f_4(t),\qquad \phi_2=f_5(x)+f_6(y)+f_7(t),\eqn{lindeteqns}$$
where the $f_i$ are arbitrary functions of their arguments.
The interpretation of this symmetry group is, that if $u=U(x,y,t)$,
$v=V(x,y,t)$ is a solution of (\eqgsg) then another solution is
$u(x,y,t)=U(f_1(x),f_2(y),f_3(t))+f_4(t)$ and
$v(x,y,t)=V(f_1(x),f_2(y),f_3(t))+
f_5(x)+f_6(y)+f_7(t)$. We shall use this classical symmetry group to
simplify the presentation of the nonclassical reduction solutions
obtained
below.

There are two canonical, classical, symmetry reductions.
\smallskip\noindent\underbar{\sl Reduction 2.1} $\xi_1 =-\mu$,
$\xi_2=1$,
$\xi_3=\phi_1=\phi_2=0$. In this case we obtain the symmetry reduction
$$u(x,y,t) = P(z,t),\qquad v(x,y,t) = Q(z,t),\qquad
z=x+\mu y,$$
where $P(z,t)$ and $Q(z,t)$ satisfy
$$P_{zzt} + 2P_zQ_{zt} = 0,\qquad Q_{zz} = P_z^2.\eqn{eqcrii}$$
This system can be viewed as a form of the real, unpumped Maxwell-Bloch
system; see equation (4.1) below with $c=0$, which arises in nonlinear
optics (cf.,
Ablowitz \&\ Segur 1981). Further, making the transformation
$P=\tfr12\theta$, $Q_{zz}=\tfr14\theta_z^2$ and $Q_{zt}=-\tfr12
\cos\theta$ in
(\eqcrii) yields the sine-Gordon equation (\eqSG). We remark that since
the system
(\eqgsg) is invariant under the transformation $(x,y)\to(f(x),g(y))$,
where
$f(x)$ and $g(y)$ are arbitrary functions, then we can generate a
variety of other,
equivalent, reductions.

\smallskip\noindent\underbar{\sl Reduction 2.2}
$\xi_1 =\kappa_1$, $\xi_2=\kappa_2$, $\xi_3=1$, $\phi_1=\phi_2=0$. In
this case
we obtain the symmetry reduction
$$u(x,y,t) = P(z,\zeta),\qquad v(x,y,t) = Q(z,\zeta),\qquad
z=x-\kappa_1t,\qquad \zeta=y-\kappa_2t,$$
where $P(z,\zeta)$ and $Q(z,\zeta)$ satisfy
$$\eqalignno{&\kappa_1P_{zz\zeta} + \kappa_2P_{z\zeta\zeta} +
P_zP_\zeta(\kappa_1P_z+\kappa_2P_\zeta) + \kappa_1P_{\zeta} Q_{zz} +
\kappa_2P_z Q_{\zeta\zeta} = 0,&\eqnm{eqcri}{a}\cr
&Q_{z\zeta} = P_zP_\zeta.&\eqnr{b}\cr}$$

\section{Nonclassical Method}
Bluman \&\ Cole (1969), in their study of symmetry reductions of
the linear heat equation, proposed the so-called ``nonclassical method
of
group-invariant solutions''; this technique is also known as the
``method of
conditional symmetries" (cf., Levi \&\ Winternitz 1989).
This method involves considerably more algebra and
associated calculations than the classical Lie method. In fact, it has
been
suggested that for some \pdes, the calculation of these nonclassical
reductions
might be too difficult to do explicitly, especially if attempted
manually,
since the associated determining equations are now an overdetermined,
{\it
nonlinear\/} system. Further, the associated vector fields arising from
the
nonclassical method do not form a vector space, still less a Lie
algebra, since
the invariant surface conditions (\insc) depend upon the particular
reduction.

In the nonclassical method one requires only the subset of ${\cal
S}_{\DELTA}$ given by
$${\cal S}_{\DELTA,\PSI} =\left\{u(x,y,t),v(x,y,t)\;:\;\Delta_1(u,v)
=0,\;\Delta_2(u,v) =0,\;\psi_1(u,v)
=0,\;\psi_2(u,v)=0\right\},\eqn{nclset}$$
is invariant under the trans\-forma\-tion ({\eqinftr}), where
${\cal S}_{\DELTA}$ is as defined in (\Sdelta) and $\psi_1
=0$ and $\psi_2=0$ are the invariant surface conditions (\insc).

As we showed in Clarkson \&\ Mansfield (1994{\it c}), the standard
procedure for
applying the nonclassical method (e.g., as described by Levi
\&\ Winternitz 1989),
can create difficulties, particularly when implemented in symbolic
manipulation
programs. These difficulties often arise for equations such as (\eqgsg)
which
require the use of differential consequences of the invariant surface
conditions
(\insc). In Clarkson \&\ Mansfield (1994{\it c}), we proposed an
algorithm for
calculating the determining equations associated with the nonclassical
method which
avoids these difficulties, and we use that algorithm here.
Further we showed how the MACSYMA package {\tt symmgrp.max}, which was
written
to calculate the determining equations for the classical method, can be
adapted
to calculate the determining equations for the nonclassical method and
we use this
modification here.

To apply the nonclassical method to the system (\eqgsg),
there are three cases to consider: (i), $\xi_3\not=0$, (ii), $\xi_3=0$
and
$\xi_2\not=0$, and (iii), $\xi_3=0$ and $\xi_2=0$ and $\xi_1\not=0$.

\subsection{Case (i). $\xi_3\not=0$}
In this case we set $\xi_3=1$, without loss of generality, and so the
invariant
surface conditions (\insc) simplify to
$$ \xi_1 u_x + \xi_2u_y +u_t=\phi_1(x,y,t,u,v), \qquad \xi_1 v_x
+\xi_2v_y
+v_t=\phi_2(x,y,t,u,v).\eqn{inscnci}$$
We first eliminate $u_{xyt}$, $v_{xt}$ and $v_{yt}$ from (\eqgsg) using
(\inscnci)
and then apply the classical Lie algorithm to the resulting equations.
This
procedure yields a system of twenty-eight determining equations, of
which twenty
were linear and eight nonlinear. Then using the MAPLE package {\tt
diffgrob2}
(Mansfield 1993) these simplify to seventeen equations, thirteen of
which are
one-term equations and the other four are two-term nonlinear equations.
This reduced
system is easily solved to yield the classical infinitesimals
(\lindeteqns). Thus,
in this generic case, the nonclassical method does not yield any
additional
symmetry reductions to those obtained using the classical Lie method.

\subsection{Case (ii). $\xi_3=0$, $\xi_2\not=0$}
In this case we set $\xi_2=1$, without loss of generality, and,
analogous to the
procedure in the generic case, we use
(\insc), with $\xi_3=0$ and $\xi_2=1$, to eliminate all
$y$-derivatives, i.e.\
$u_{xyt}$, $u_y$, $v_{yt}$ and $v_{xy}$, from (\eqgsg). Then applying
the
classical Lie algorithm to the resulting equations yields the following
determining equations,
$$\eqalignno{
\xi_{1,v}=0,\qquad \xi_{1,u}=0,\qquad \xi_{1,t}=0,\qquad
\xi_{1,x} \xi_{1,y} - \xi_1 \xi_{1,xy}&=0,&\eqnm{nciideteqns}{a}\cr
\phi_{1,u}=0,\qquad \phi_{1,v}=0,\qquad
\phi_{2,u}=0,\qquad \phi_{2,v}&=0,&\eqnr{b}\cr
\xi_1 \phi_{1,y} - \xi_1^{2} \phi_{1,x} - \xi_{1,y}
\phi_1&=0,&\eqnr{c}\cr
\xi_1 \phi_{2,yt} - \xi_1^{2} \phi_{2,xt} - \xi_{1,y}
\phi_{2,t}&=0,&\eqnr{d}\cr
\xi_1 \phi_{2,xy} - \xi_{1,y} \phi_{2,x} - \xi_1 \phi_1
\phi_{1,x}&=0,&\eqnr{e}\cr
\xi_1\phi_1 \phi_{2,xt} + \xi_1\phi_{1,x} \phi_{2,t}
 + \xi_{1} \phi_{1,xyt} -\xi_{1,y}\phi_{1,xt}&=0.&\eqnr{f}\cr}$$
The system (\nciideteqns) is straightforward to analyse.
{}From (\nciideteqns{\it a,b,c,d}), one obtains that,
$$\xi_1={g_2'(y)/g_1'(x)},\qquad
\phi_1=g_2'(y)w_1(z,t),\qquad
\phi_2=g_2'(y)\left[w_2(z,t)+w_3(x,y)\right],$$
where $z=g_1(x)+g_2(y)$, with $g_1(x)$ and $g_2(y)$ arbitrary
functions.
Substituting these into (\nciideteqns{\it e,f}) it is obtained that,
$$\eqalignno{
w_1w_{2,t}+w_{1,zt}+m_1(t)&=0,&\eqnm{eqws}{a}\cr
2w_{2,z}-w_1^2+m_2(t)&=0,&\eqnr{b}\cr
w_3+H(x)+K(y)&=0,&\eqnr{c}\cr}$$
where $m_1(t)$, $m_2(t)$, $H(x)$ and $K(y)$ are arbitrary
functions of integration (we have set a third arbitrary function of $z$
to zero,
without loss of generality). It transpires that after the solutions $u$
and
$v$ have been obtained, we can use the classical symmetries
(\lindeteqns) to set
$g_1(x)=x$, $g_2(y)=y$, $H(x)=0$ and $K(y)=0$.
Thus, we do this now, to keep the exposition simple.
Putting everything together, we
obtain the following solution to the determining equations,
$$\xi_1=1,\qquad
\xi_2=1,\qquad
\xi_3=0,\qquad
\phi_1=w_1(z,t),\qquad
\phi_2=w_2(z,t),\eqn{ncinfii}$$
where $z=x+y$ and $w_1(z,t)$ and $w_2(z,t)$ satisfy (\eqws{\it a,b}).

We digress to remark that eliminating $w_2$ from (\eqws{\it a,b}) and
setting
$w_1=w$ yields,
$$ww_{zzt}- w_zw_{zt}-m_1(t) w_{z}
+w^3w_{t}-\tfr12{\d m_2\over\d t}w^2=0.\eqn{eqwI}$$
In \S6.2 below we apply the \p\ test due to Weiss \etal\ (1993) and
discuss the
integrability of this equation. We note that in the special case when
$\displaystyle
m_1(t)\propto {\d m_2\over\d t}$, we can simplify this equation by
rescaling $t$. In
this case, it is sufficient to set $m_1(t)=\mu_1$ and $m_2(t)=\mu_2t$,
with $\mu_1$
and $\mu_2$ constants, without loss of generality and so we obtain,
$$ww_{zzt}- w_zw_{zt} -\mu_1 w_{z}
+w^3w_{t}-\tfr12\mu_2w^2=0.\eqn{eqwII}$$

Next, we set $w_1(z,t)=2P_{z}(z,t)$ and $w_2(z,t)=2Q_{z}(z,t)$ in
(\ncinfii) and
then integrate the invariant surface conditions (\insc) using the
method of
characteristics. Hence we obtain the nonclassical reduction of (\eqgsg)
given by
$$u(x,y,t)=P(z,t)+ \~P(\tz,t),\qquad
v(x,y,t)=Q(z,t)+ \~Q(\tz,t),\eqn{uvint}$$
where $z=x+y$ and $\tz=x-y$,
$P(z,t)$ and $Q(z,t)$ satisfy
$$\eqalignno{
P_{zzt}+2P_{z}Q_{zt}+\tfr12m_1(t)&=0,&\eqnm{eqXXW}{a}\cr
Q_{zz}-P_{z}^2+\tfr14m_2(t)&=0,&\eqnr{b}\cr}$$
$\~P(\tz,t)$ and $\~Q(\tz,t)$ satisfy the same equations,
and $m_1(t)$ and $m_2(t)$ are arbitrary functions, which
are effectively separation functions.

In \S4 below, we obtain exact solutions of the sine-Gordon system
(\eqgsg), by
analysing the reduced system (\eqXXW), or equivalently (\eqwI), for
some special
choices of $m_1(t)$ and $m_2(t)$.

\subsection{Case (iii). $\xi_3=0$, $\xi_2=0$, $\xi_1\not=0$}
In this case we set $\xi_1=1$, without loss of generality, and, as
above, we use
(\insc), with $\xi_3=0$, $\xi_2=0$ and $\xi_1$, to eliminate all
$x$-derivatives,
i.e.\ $u_{xyt}$, $u_x$, $v_{xt}$ and $v_{xy}$, from (\eqgsg).
Then applying the classical Lie algorithm to the resulting equations
yields seven
equations containing over twelve hundred terms. Since this system is
considerably
more complex than the sine-Gordon system (\eqgsg) we are studying,
we shall not pursue this case further here.

\subsection{Nonclassical reductions of (\eqKR) and (\eqKRa)}
In this subsection, we record, for the sake of completeness, some
nonclassical
reductions of the $2+1$-dimensional sine-Gordon system (\eqKR) and the
$2+1$-dimensional sine-Gordon equation (\eqKRa).
By considering the gauge transformation (\eqgauge) and the nonclassical
reduction
(\uvint) of (\eqgsg) in the case when $m_1\equiv0$ and $m_2\equiv0$, it
is
straightforward to obtain the  nonclassical reduction of the
$2+1$-dimensional
sine-Gordon system (\eqKR) given by
$$\eqalignno{\Theta(X,Y,T) &= \theta(X,T) + \~\theta(Y,T),\qquad
\Phi(X,Y,T) =\phi(X,T) +\~\phi(Y,T),&\eqnn{eqKRnc}\cr}$$
where $\theta(X,T)$, $\~\theta(Y,T)$, $\phi(X,T)$ and $\~\phi(Y,T)$
satisfy
$$\eqalignno{&\theta_{XT} = F(T)\sin\theta,\qquad\phi_{XT}
= -\~F(T)\sin\phi, &\eqnm{eqKRnci}{a}\cr &\~\theta_{YT} =
\~F(T)\sin\~\theta,\qquad \~\phi_{YT} = -F(T)\sin\~\phi,&\eqnr{b}\cr}$$
where $F(T)$
and $\~F(T)$ are arbitrary functions.

The analogous nonclassical reduction of the $2+1$-dimensional
sine-Gordon equation (\eqKRa) given by
$$\Theta(X,Y,T) = \theta(X,T) + \~\theta(Y,T),\eqn{eqKRanc}$$
where $\theta(X,T)$ and $\~\theta(Y,T)$ satisfy
$$\theta_{XT} = F(T)\sin\theta,\qquad \~\theta_{YT} = -
F(T)\sin\~\theta,\eqn{eqKRanci}$$ where $F(T)$ is an arbitrary
function.

\section{Some Exact Solutions of the sine-Gordon system (\eqgsg)}
\subsection{Nonclassical reduction solutions and the Maxwell-Bloch
system}
Equations (\uvint) describe nonclassical reductions to the
$2+1$-dimensional
sine-Gordon system (\eqgsg), in terms of the system (\eqXXW), or
equivalently
(\eqws{\it a,b}). System (\eqws{\it a,b}) is easily seen to be a
generalisation
of the real, pumped, Maxwell-Bloch system,
$$E_{\chi}=\rho,\qquad\rho_\tau=E\eta,\qquad\eta_\tau+E\rho=c,\eqn{eqMBp}$$
where $c$ is a constant, which has a physical interpretation in
nonlinear optics
(Burtsev \& Gabitov 1994). Eliminating $\rho$, and setting
$\chi\equiv t$, $\tau\equiv z$, $E\equiv w_1$ and $\eta\equiv -w_{2,t}$
in
(\eqMBp) yields (\eqws{\it a,b}) with $m_1\equiv 0$ and $m_2 \equiv
2ct$.
For $c=0$, the unpumped case, the system (\eqMBp) is equivalent to the
 sine-Gordon equation (\eqSG), which is solvable by inverse scattering
(Ablowitz \etal\ 1974). However for $c\ne 0$, the system (\eqMBp) is
solvable
by inverse scattering with a non-isospectral Lax pair (Burtsev
\etal\ 1987);
see \S6.2 below. Several authors have obtained classical reductions of
(\eqMBp)
to the fifth Painlev\'e equation (see, for example Burtsev 1993; Kitaev
\etal\ 1993; Schief 1994; Winternitz 1992), and a \bt\ for (\eqMBp) is
discussed by Schief (1994).

In this section, we examine some special cases of the system (\eqXXW),
or
equivalently (\eqwI). First, in the case $m_1(t)=m_2(t)\equiv 0$, we
express
solutions of the system in terms of solutions of the sine-Gordon
equation (\eqSG).
This yields solutions of (\eqgsg)  that can be written as a decoupled
sum of
solutions of the sine-Gordon equation, with different scalings of the
time variable
in each summand being allowed. Second, in the case $\displaystyle
m_1(t)\propto {\d
m_2\over\d t}(t)$, we obtain scaling reductions and travelling wave
solutions which
can be expressed in terms of solutions of the second, third, and fifth
Painlev\'e
equations. For each of these cases, we give the necessary
trans\-forma\-tions to
obtain the associated solutions of (\eqgsg), and write down some
examples. Indeed,
hierarchies of these solutions of (\eqgsg) are obtained, with explicit
trans\-forma\-tions connecting the members of each hierarchy given.

\subsection{The case $m_1(t)=m_2(t)\equiv 0$}
First we note that in this case, the system (\eqws)
has a classical symmetry which allows an arbitrary rescaling of the
$t$ variable. As noted in \S2, the trans\-forma\-tion $P=\tfr12\theta$,
$Q_{zz}=\tfr14\theta_z^2$ and $Q_{zt}=-\tfr12 \cos\theta$ maps (\eqXXW)
with
$m_1(t)=m_2(t)=0$ into the sine-Gordon equation (\eqSG). Hence in this
case we
obtain the solution to (\eqgsg) given by
$$\eqalignno{
u(x,y,t) &= \tfr12\theta(x+y,\tau(t))+\tfr12\~\theta(x-y,\~\tau(t)),
&\eqnm{uvinti}{a}\cr
v(x,y,t) &=
\tfr14\int^{x+y}\int^{z}\theta(\zeta,\tau(t))\,\d\zeta\,\d z+
\tfr14\int^{x-y}\int^{\~z}\~\theta(\tzeta,\~\tau(t))\,\d\tzeta\,\d
\~z,&\eqnr{b}\cr}$$ where $\theta(z,\tau)$ and $\~\theta(\~z,\~\tau)$
are
any two solutions of (\eqSG). Three well-known solutions of (\eqSG) are
$$\eqalignno{
\theta(z,t)&=4\tan^{-1}\left\{C\exp\left({\beta
z+t/\beta}\right)\right\},
&\eqnm{sgkink}{}\cr
\theta(z,t)&=4\tan^{-1}\left\{{\lambda\over\mu}\sin\left(\mu z-
{\mu t\over\lambda^2+\mu^2}\right)\sech\left(\lambda z +
{\lambda t\over \lambda^2+\mu^2}\right)\right\},&\eqnm{sgbre}{}\cr
\theta(z,t)&=4\tan^{-1}\left\{\left(\beta_2+\beta_1\over
\beta_2-\beta_1\right){\exp(\eta_2)-\exp(\eta_1 )\over 1+
\exp(\eta_1 +\eta_2)}\right\},&\eqnm{sgkk}{}\cr}$$
where $\eta_1=\beta_1 z+t/\beta_1$, and $\eta_2=\beta_2 z+t/\beta_2$.
The first of these is the one-soliton, or kink, solution,
the second (\sgbre), is the ``breather" solution, while the third,
(\sgkk), is the kink-kink solution (if $\beta_2>-\beta_1>0$)
or kink-antikink solution (if $\beta_2>\beta_1>0$).
In Figure 1 we plot (\uvinti{\it a}) when $\theta(z,\tau)$ and
$\~\theta(\~z,\~\tau)$ are both breathers and in Figure 2 when
$\theta(z,\tau)$
and $\~\theta(\~z,\~\tau)$ are both kink-kink solutions. Since one can
rescale
$x$, $y$ and $t$ arbitrarily, using the classical symmetries of
(\eqgsg), a wide
variety of solutions is obtainable. This is illustrated in Figure 3
where
$\theta(z,\tau)$ and $\~\theta(\~z,\~\tau)$ are both breather solutions
but
$x$ has been scaled to $-\exp(1-x^2)$, $\tau(t)=\exp(1-t^2)$ and
$\~\tau(t)=-\exp(1-t^2)$. This last figure illustrates the fundamental
difficulties
of attempting to solve the system (\eqgsg) numerically. In particular,
one would not
be able to distinguish the solutions in an initial value problem since
an
exponentially small change in the initial conditions can result in
completely
different qualitative behaviours.

\subsection{The case $\displaystyle m_1(t)\propto {\d m_2\over\d
t}(t)$}
To obtain exact solutions in this case, it is sufficient to
analyse (\eqwII), which has a three-dimensional
group of classical Lie point symmetries, given by the vector fields
$${\bf v}_1=\pa_z,\qquad{\bf v}_2=\pa_{t},\qquad {\bf v}_3=2t\pa_{t}
-z\pa_z+w\pa_w.$$

The first two vector fields yield the travelling wave reduction
$$w(z,t)=\eta(\zeta),\qquad \zeta=z-\beta t,\eqn{eqXXXtw}$$
where $\eta(\zeta)$ satisfies
$$\eta_{\zeta\zeta}=-\tfr12\eta^3-\mu_2{\zeta +\cc1\over 2\beta} \eta+
{\mu_1\over \beta},\eqn{wPII}$$ where $\kappa_1$ is a constant of
integration.
Equation (\wPII) is equivalent to second Painlev\'e equation (\pii)
$$ y_{xx}=2y^3+xy+\alpha\eqn{eqPII}$$
(cf., Ince 1956), provided that $\mu_2\not=0$; if $\mu_2=0$ then
(\wPII) is solvable
in terms of Jacobi elliptic functions (cf., Whittaker \&\ Watson
1927).

The third vector field ${\bf v}_3$ yields the scaling reduction
$$w(z,t)=z^{-1} F(\zeta),\qquad\zeta=z{t}^{1/2},$$
where $F(\zeta)$ satisfies
$$\zeta^2F{\Fzzz}-\zeta^2{\Fz}{\Fzz}+\zeta F{\Fzz}+
F^3{\Fz}+2\mu_1\left(\zeta F -
\zeta^2{\Fz}\right)-\mu_2\zeta F^2=0.\eqn{scalredi}$$
In the case $\mu_1=0$, this is exactly equation (5.5) of Winternitz
(1992),
with $c=\tfr18$. Equation (\scalredi) has a first integral
of the form,
$$\left({\Fzz}-\mu_2F+2\mu_1\right)^2\zeta^{2}
+F^2\left\{\left({\Fz}\right)^2-{\mu_2}F^2
+4\mu_1F-K\right\}=0. \eqn{scalred}$$%
We note here that this integral appears not to be obtainable by either
the
classical or contact symmetry methods.
Following the analysis in Bureau (1972), p.\ 212, we can write the
solutions of (\scalred) in terms of the third and fifth Painlev\'e
equations.
It transpires that there are three cases to consider, (i) $\mu_1$
arbitrary and
$\mu_2 \ne 0$, (ii) $\mu_1\ne 0$ and $\mu_2=0$, and (iii)
$\mu_1=\mu_2=0$.

\smallskip\noindent{\bf Case \secsub(i)}. $\mu_1$ arbitrary and $\mu_2
\ne 0$.
In this case, the trans\-forma\-tion to be used in (\scalred), as
given by Bureau's method, is
$$F(\zeta)={\i\over\psi(\xi)[\psi(\xi)-1]}
\left\{2\xi{\psi_{\xi}(\xi)}+1-\psi(\xi)-
{2\i \mu_1\over\mu_2}[\psi(\xi) -1]^2\right\},
\qquad \xi=\tfr12\zeta^2, \eqn{FPVsub}$$
where $\psi(\xi)$ satisfies the fifth Painlev\'e equation (\pv),
$${\psi_{\xi\xi}}=\left({1\over 2\psi}+{1\over
\psi-1}\right)\psi^2_{\xi}-{1\over
\xi}{\psi_{\xi}}+{(\psi-1)^2\over
\xi^2}\left(\alpha_5\psi+{\beta_5\over
\psi}\right)+{\gamma_5\over \xi}\psi +{\delta_5\psi(\psi+1)\over
\psi-1},\eqn{eqPV}$$
and the constants are given by
$$\alpha_5={(K\mu_2-4\mu_1^2)/(8\mu_2^2)},\qquad
\beta_5=-{(\mu_2-2\i\mu_1)^2\!/(8\mu_2^2)},\qquad
\gamma_5=-\tfr14{\mu_2},\qquad
\delta_5=0.$$

It is known that the special case of \pv\ with $\delta_5=0$ can always
be
solved in terms of solutions of the third Painlev\'e equation (\piii)
$${\phi_{\zeta\zeta}}={1\over \phi}\left({\phi_{\zeta}}\right)^2
-{1\over \zeta}{\phi_{\zeta}}+{\alpha_3 \phi^2 +\beta_3\over
\zeta}+\gamma_3\phi^3+{\delta_3\over\phi}\eqn{eqPIII}$$
(cf., Fokas \&\ Ablowitz 1982; Gromak 1975);
Cosgrove \&\ Scoufis (1993), remark that there are infinitely many
other special
cases of \pv\ that are solvable in terms of solutions of \piii.
Combining (\FPVsub)
with the trans\-forma\-tion of \pv\ to \piii\ given in Gromak (1975),
one
obtains the trans\-forma\-tion to be used in (\scalred) to be
$$F(\zeta)=-{\i\zeta\left\{{\phi_{\zeta}(\zeta)}+c\phi^2(\zeta)+
d\right\}\!/\phi(\zeta)},$$ where $\phi(\zeta)$ satisfies
\piii\ (\eqPIII) with
constants given by
$$\alpha_3={\sqrt{K\mu_2-4\mu_1^2}+\mu_2-2\i\mu_1\over4d},
\quad \beta_3={\sqrt{K\mu_2-4\mu_1^2}+\mu_2+2\i\mu_1\over4c},
\quad \gamma_3=c^2,\quad\delta_3=-d^2,\eqn{eqPIIIi}$$ with
$\mu_2=-4cd$.

\smallskip\noindent{\bf Case \secsub(ii)}. $\mu_1\ne 0$ and $\mu_2=0$.
In this case, the
trans\-forma\-tion used is
$$F(\zeta)={\i\zeta}[{\phi_{\zeta}(\zeta)}-\tfr12\nu]/\phi(\zeta),$$
where $\nu$ is a non-zero constant and
$\phi(\zeta)$ satisfies \piii\ (\eqPIII) with constants
$\alpha_3=2\i{\mu_1 /\nu}$, $\beta_3=\tfr18{\nu}\left(4+{\i
K/\mu_1}\right)$, $\gamma_3=0$ and $\delta_3=-\tfr14\nu^2$.

\smallskip\noindent{\bf Case \secsub(iii)}. $\mu_1 =\mu_2=0$. In this
case, the
trans\-forma\-tion used is
$F(\zeta)={\i\zeta}\phi_{\zeta}(\zeta)/\phi(\zeta)$,
where
$\phi(\zeta)$ satisfies \piii\ (\eqPIII) with constants
$\alpha_3=1$, $\beta_3=-\tfr14K$ and $\gamma_3=\delta_3=0$.

\smallskip In the following two subsections, we write down the
solutions to
(\eqgsg) in terms of solutions of \pii\ and \piii. Using some known
rational and other solutions of these \p\ equations, we obtain some
simple
explicit solutions for the sine-Gordon system (\eqgsg). Further, we
obtain
\bts\ for each subfamily of solutions.

\subsection{Nonclassical solutions in terms of the second \p\ equation}
Here we consider travelling wave reductions to (\eqwII) given by
(\eqXXXtw),
to obtain solutions to (\eqgsg) in terms of \pii\ (\eqPII).
We set $\mu_2=1$ in (\wPII) for convenience. Following the method of
integration in
\S3.2, we obtain the solution to (\eqgsg), after a complex rescaling of
$\eta(\zeta)$ and simplifying using the classical symmetry group,
$$\eqalignno{
u(x,y,t)&=\i\left\{ \int^{\zeta}\eta(\zeta_1;\mu_1,1,\beta)\,\d\zeta_1+
\int^{\tzeta}\~
\eta(\tzeta_1;\mu_1,1,\~\beta)\,\d\tzeta_1\right\},&\eqnm{uvextra}{a}\cr
v(x,y,t)&=-\int^{\zeta}\!\!\!\int^{\zeta_1}[\eta(\zeta_2;\mu_1,1,\beta)]^2
\,\d\zeta_2\,\d\zeta_1
-\int^{\tzeta}\!\!\!\int^{\tzeta_1}[\~\eta(\tzeta_2;\mu_1,1,\~\beta)]^2
\,\d\tzeta_2\,\d\tzeta_1 \cr&\qquad
-\tfr14(x^2+y^2)t+\tfr18[\beta(x+y)+\~\beta(x-y)]t^2
-\tfr14[\cc1(x+y)+\cc2(x-y)]t,{\hbox to 20pt{\hfill}}&\eqnr{b}\cr}$$
where
$\eta(\zeta;\mu_1,\mu_2,\beta)$ and
$\~\eta(\tzeta;\mu_1,\mu_2,\~\beta)$ satisfy
(\wPII). It is well-known that \pii\ (\eqPII) possesses rational
solutions and one-parameter families of solutions expressible in terms
of Airy
functions for special choices of the parameter $\alpha$
(cf., Airault 1979; Clarkson 1990; Gibbon \etal\ 1988; Murato, 1985).
For example, the first three non-zero rational solutions $y(x;\alpha)$
of \pii\
(\eqPII) are given by
$$ y(x;1) = -\,{1\over x},\qquad y(x;2) = -\,{2(x^3-2)\over
x(x^3+4)},\qquad y(x;3) = -\,{3x^2(x^6+8x^3+160)\over
(x^3+4)(x^6+20x^3-80)}.$$
Thus, we obtain the simple solution of (\eqgsg),
$$\eqalignno{
u_1(x,y,t)&=\i\ln(\zeta\tzeta),&\eqnm{gsgsoli}{a}\cr
v_1(x,y,t)&=\ln(\zeta\tzeta)-\tfr14(x^2+y^2)t-\tfr18[
c(x+y)+\beta(x-y)]t^2,&\eqnr{b}\cr}$$
while setting $c=\beta=-\tfr12$, and then using the classical
symmetry to rescale $t$ (we let $t$ go to $2t$ for convenience) so that
$\zeta=x+y+t$ and $\tzeta=x-y+t$, we obtain,
$$\eqalignno{
u_2(x,y,t)&=\i\left\{\ln(\zeta\tzeta)-
\ln\left[(\zeta^3+4)(\tzeta^3+4)\right]\right\},
&\eqnm{gsgsolii}{a}\cr
v_2(x,y,t)&=\ln[\zeta\tzeta(\zeta^3+4)(\tzeta^3+4)]
-\tfr14(x^2+y^2)t-\tfr12xt^2,&\eqnr{b}\cr\cr
u_3(x,y,t)&=\i\left\{\ln\left[(\zeta^3+4)(\tzeta^3+4)\right]-
\ln\left[(\zeta^6+20\zeta^3-80)(\tzeta^6+20\tzeta^3-80)\right]\right\},
&\eqnm{gsgsoliii}{a}\cr
v_3(x,y,t)&=\ln[(\zeta^3+4)(\tzeta^3+4)(\zeta^6+20\zeta^3-80)
(\tzeta^6+20\tzeta^3-80)]
-\tfr12(x^2+y^2)t-\tfr12xt^2.{\hbox to 30pt{\hfill}}&\eqnr{b}\cr}
$$

It can be seen that these solutions are not
obtainable from simple solutions using the \bt\ (5.2)
for the system given in \S6 below. However, one may use the
well-known \bt\ for \pii\ (\eqPII),
$$ y(x;\alpha+1)=-y(x;\alpha)-{1+2\alpha\over
2y^2(x;\alpha)+2y_x(x;\alpha)+x}$$
(cf., Airault 1979), to obtain
trans\-forma\-tions within the family of solutions (\uvextra).
It is straightforward to show that
$$\eqalign{
\int^x y(x_1;\alpha+1)\,\d x_1&=
\int^x y(x_1;\alpha)\,\d x_1
-\ln\left\{2[y(x;\alpha)]^2+2y_x(x;\alpha)+x\right\},\cr
\int^x\!\int^{x_1} [y(x_2;\alpha+1)]^2\,\d x_2\d x_1&=-
\int^x\!\int^{x_1}[y(x_2;\alpha)]^2\,\d x_2\d x_1
+\ln\left\{2[y(x;\alpha)]^2+2y_x(x;\alpha)+x\right\}.\cr}$$
Thus from (\uvextra),
$$\eqalign{
\~u(x,y,t)= u(x,y,t)&-
\i\ln\left[2\eta^2(\zeta;\mu_1,\mu_2,\beta)+
2\eta_{\zeta}(\zeta;\mu_1,\mu_2,\beta)+\zeta\right]
\cr&-\i\ln\left[2\~\eta^2(\tzeta;\mu_1,\mu_2,\~\beta)+
2\~\eta_{\tzeta}(\tzeta;\mu_1,\mu_2,\~\beta)+\tzeta\right],\cr
\~v(x,y,t)= v(x,y,t)&+2\i u(x,y,t)+
\ln\left[2\eta^2(\zeta;\mu_1,\mu_2,\beta)+
2\eta_{\zeta}(\zeta;\mu_1,\mu_2,\beta)+\zeta\right]
\cr&+\ln\left[2\~\eta^2(\tzeta;\mu_1,\mu_2,\~\beta)+
2\~\eta_{\tzeta}(\tzeta;\mu_1,\mu_2,\~\beta)+\tzeta\right],\cr}$$
is the \bt\ acting within the family of solutions (\uvextra).

\subsection{Nonclassical solutions in terms of the third \p\ equation}
The relationship between solutions of the ordinary differential
equation (\scalred)
to solutions of the sine-Gordon  system (\eqgsg) is given by the
following. If
$F(\zeta)$ and $\^F(\zeta)$ both satisfy (\scalred), then the
associated solution of
(\eqgsg) is given by,
$$\eqalignno{u(x,y,t)&=
\tfr12\int^{zt^{1/2}} {F(\zeta_1)\over \zeta_1}\,\d \zeta_1+
\tfr12\int^{\^zt^{1/2}}{\^F(\^\zeta_1)\over \^\zeta_1}\,\d
\^\zeta_1&\eqnm{uvscal}{a}\cr v(x,y,t)&=
\tfr14\!\int^{zt^{1/2}}\!\!\int^{\zeta_1}\!\!\left[{F(\zeta_2)\over
\zeta_2}\right]^2
\d \zeta_2 \d \zeta_1\!
+\!\tfr14\int^{\^zt^{1/2}}\!\!\int^{\^\zeta_1}\!\!\left[{\^F(\^\zeta_2)\over
\^\zeta_2}\right]^2
\d \^\zeta_2 \d \^\zeta_1 -\tfr18\mu_2 t(z^2+{\^z}^2),{\hbox to
20pt{\hfill}}
&\eqnr{b}}$$
where $z=x+y$ and $\^z=x-y$. In \S4.3 above, we noted the relationship
between the
equation (\scalred) and \piii\ (\eqPIII). There are many known rational
and
one-parameter family solutions expressible in terms of Bessel functions
for \piii\
(cf., Lukashevich 1965; Gromak 1977; Okamoto 1987), and \bts\ which map
solutions of
\piii\ into new solutions with different values of the parameters (cf.,
Airault
1979; Fokas \&\ Ablowitz 1982; Milne \&\ Clarkson 1993). In the
following, we obtain
examples of solutions for (\eqgsg) associated with solutions of \piii.
Further, we
discuss the relationship between the equation (\scalred) and the third
\p\ equation
(\eqPIII), in more detail. The result will be hierarchies of solutions
for the
system (\eqgsg), and trans\-forma\-tions between solutions in each
hierarchy.

\smallskip\noindent{\bf Case \secsub(i)} $\mu_1$ arbitrary, $\mu_2\ne
0$.\quad
The equations (\scalred) and (\eqPIII) are related by the system,
$$\eqalignno{
\left[F_\zeta+2\i \phi (\i c
F+c+\alpha_3)\right]^2+F^2\mu_2-4F\mu_1+K-F_\zeta^2&=0,&
\eqnm{Fysys}{a}\cr
\phi F+\i \zeta(\phi _\zeta+c \phi^2+d)&=0,&\eqnr{b}\cr}$$
where
$\mu_1=\i(\beta_3 c-\alpha_3 d)$, $\mu_2=-4cd$ and
$K=-4(\alpha_3+c)(\beta_3+d)$,
with $\gamma_3=c^2$ and $\delta_3=-d^2$. Hence we obtain the following
result.

\def\Mu{\kappa}
\medskip\vbox{{\bf Theorem \secsub.1}. \par
(i)\enskip{\it  If $F(\zeta;\mu_1,\mu_2,K)$ is a solution of
$(\scalred)$ then
$$\phi (\zeta;\alpha_3,\beta_3,\gamma_3,\delta_3)={\i
\mu_2\left\{F_\zeta(\zeta)\mp
[F_\zeta^2(\zeta)-\mu_2F^2(\zeta) +4\mu_1F(\zeta)-K]^{1/2}\right\}
\over
2\epsilon_1 [\Mu+\i \mu_2F(\zeta)-2\i\mu_1]},\eqn{FBTy}$$
where $\Mu =\pm(\mu_2K-4\mu_1^2)^{1/2}$,
is a solution of the \piii\ $(\eqPIII)$ with parameters
$$\alpha_3={\epsilon_1(\Mu-\mu_2-2\i\mu_1)/\mu_2},\qquad
\beta_3=-{(\Mu-\mu_2+2\i\mu_1)/(4\epsilon_1)},\qquad
\gamma_3=\epsilon_1^2,\qquad\delta_3 =-{\mu_2^2/(16\epsilon_1^2)}.$$}}

(ii)\enskip{\it  If $\phi (\zeta;\alpha_3,\beta_3,\gamma_3,\delta_3)$
is a solution of
\piii\ $(\eqPIII)$ with $\gamma_3=\epsilon_3^2$ and
$\delta_3=-\epsilon_4^2,$ then
$$F(\zeta;\mu_1,\mu_2,K)=-{\i \zeta[\phi_\zeta(\zeta)+\epsilon_3
\phi^2(\zeta)+\epsilon_4]/\phi(\zeta) }\eqn{yBTF}$$ is a solution of
$(\scalred)$
with parameters
$\mu_1=\i(\beta_3\epsilon_3-\alpha_3\epsilon_4)$,
$\mu_2=-4\epsilon_3\epsilon_4$,
$K=-4(\alpha_3+\epsilon_3)(\beta_3+\epsilon_4)$.}
\smallskip

Setting $\epsilon_2=-\mu_2/(4\epsilon_1)$, and
$\epsilon_1^2=\epsilon_3^2$ and $\epsilon_2^2=\epsilon_4^2$, then
combining (\yBTF) and (\FBTy) yields the
following \bt\ for the equation (\scalred).

\medskip{\bf Theorem \secsub.2}. {\it If $F(\zeta;\mu_1,\mu_2,K)$ is a
solution of
$(\scalred)$, then
$$\^F(\zeta;\^\mu_1,\^\mu_2,\^K)= -{\zeta\over \mu_2R}[2\i
\epsilon_1\mu_2R_\zeta(\Mu+\i
\mu_2F-2\i\mu_1)+4\epsilon_1^2\epsilon_4(\Mu+\i\mu_2 F-2\i\mu_1)^2+
2\epsilon_1\mu_2^2F_\zeta R -\epsilon_3\mu_2^2R^2],\eqn{FBTscal}$$
where
$$R(\zeta)=F_\zeta(\zeta)\mp
[F_\zeta^2(\zeta)-\mu_2F^2(\zeta)+4\mu_1F(\zeta)-K]^{1/2},\qquad
\Mu=\pm(\mu_2K-4\mu_1^2)^{1/2},$$
is also a solution of $(\scalred)$ with parameters
$$\eqalign{&\^\mu_1=
-[\i(\epsilon_3\mu_2+4\epsilon_1^2\epsilon_4)(\Mu-\mu_2)
+2(4\epsilon_1^2\epsilon_4
-\epsilon_3\mu_2)\mu_1]/(4\epsilon_1\mu_2),\qquad\^\mu_2=-4\epsilon_3\epsilon_4,\cr
&\^K={1\over\epsilon_1\mu_2}[(\epsilon_3\mu_2-2\epsilon_1\mu_2
-4\epsilon_1^2\epsilon_4)\Mu+(\epsilon_1-\epsilon_3)\mu_2^2
+(K-4\epsilon_3\epsilon_4
+4\epsilon_1\epsilon_4)\epsilon_1\mu_2+2\i\mu_1(\epsilon_3\mu_2+
4\epsilon_1^2\epsilon_4)].}$$}

As an example, we take the simple ``seed solution" for \piii\ given by
$\phi(\zeta;a,-a,1,1)=1$. Applying the trans\-forma\-tion (\yBTF) to
this solution
yields the solution of (\scalred) given by
$$F(\zeta;-2\i a,-4,4(a^2-1))=-2\i\zeta.\eqn{seeda}$$
Applying the \bt\ (\FBTscal) to (\seeda) taking the upper sign
and with $\epsilon_1=\epsilon_2=1$, $\epsilon_3=\epsilon_4=-1$ yields
$$F(\zeta;2\i a,-4,4(a^2-9))= {{2\i \zeta \left(4
\zeta^{2}+4 a \zeta+a^{2}+3\right)}\over{\left(2 \zeta+a-1
\right) \left(2 \zeta+a+1\right)}}.\eqn{seedb}$$
Alternatively, applying the \bt\ (\FBTscal) to the solution (\seeda)
with
$\epsilon_1=1$, $\epsilon_2=1$, $\epsilon_3=-1$, $\epsilon_4=1$ yields
the solution
to (\scalred) given by
$$F(\zeta;4\i,-4,4(a-3)(a+1))= {4\i\zeta/(2\zeta + a -1)}.$$

It is known that \piii\ (\eqPIII) possesses a one-parameter family of
solutions
characterised by the Riccati equation
$$\phi_\zeta=-\gamma_3^{1/2}\phi^2 +
(1+\alpha_3\gamma_3^{-1/2})\phi/\zeta-(-\delta_3)^{-1/2},\eqn{eqPIIRic}$$
provided that the parameters satisfy the restriction
$2+\alpha_3\gamma_3^{-1/2}+\beta_3 (-\delta_3)^{-1/2}=0$ (Lukashevich
1965; Gromak
1977). Hence there exist solutions of \piii\ (\eqPIII) of the form
$\phi (\zeta)=\gamma_3^{-1/2}\Phi_\zeta(\zeta)/\Phi(\zeta)$,
where $\Phi(\zeta)$ satisfies
$$\Phi_{\zeta\zeta}+(1+\alpha_3\gamma_3^{-1/2})\zeta^{-1}\Phi_\zeta
+\gamma_3^{1/2}(-\delta_3)^{-1/2}\Phi = 0.$$
This equation is solvable in terms of Bessel functions or modified
Bessel functions
depending on the signs of $\gamma_3^{1/2}$ and $(-\delta_3)^{1/2}$.
There are
essentially four cases, of which we discuss two next.

\piii\ possesses the one-parameter family solution,
$$\phi_\pm(\zeta;a-1,a-1,1,-1)
={\d\over\d\zeta}\left\{\ln\left[A\zeta^{-\nu}
J_{\nu}(\pm\zeta)+B\zeta^{-\nu}Y_{\nu}(\pm\zeta)\right]\right\},
\eqn{piiiJYsol}$$
where $\nu=\tfr12(a-1)$, $J_{\nu}(\zeta)$ and $Y_{\nu}(\zeta)$ are
Bessel functions
of the first and second kind, with $A$ and $B$ arbitrary constants.
Applying the
trans\-formation (\yBTF) with $\epsilon_3=\epsilon_4=-1$, to
(\piiiJYsol),
yields the solution of (\scalred) given by
$$F_0(\zeta;2\i
a,-4,4(a^2-4))=\i{[2\zeta\phi_\pm^2(\zeta)+a\phi_\pm(\zeta)
+2\zeta]/\phi_\pm(\zeta)}.\eqn{FBaseed}$$
We now apply four different versions of the \bt\ (\FBTscal) to
the solution (\FBaseed), taking the lower sign to obtain further
solutions
rational in both $\zeta$ and $\phi_-(\zeta)$, the logarithmic
derivative of
$\Phi$. This yields the same solution, if
$\epsilon_1=\epsilon_2=\epsilon_3=\epsilon_4=1$, and
$$\eqalign{F_1&(\zeta;-2\i a,-4,4(a^2-16))\cr&=\i{(a+4)\zeta^2\phi_-^4
+(2a^2+4a-4)\zeta\phi_-^3
+[2a\zeta^2+(a^3-a)]\phi_-^2
+(2a^2-4a-4)\zeta\phi_-
+(a-4)\zeta^2\over [\zeta\phi_-^2+(a-1)\phi_-
+\zeta][\zeta\phi_-^2+(a +1)\phi_- +\zeta]},\cr
F_2&(\zeta;6\i,4,4(a-2)(a+4))=\i{(a+4)\zeta\phi_-^3+(2\zeta^2+a^2+a)
\phi_-^2
+(3a+2)\zeta\phi_-+2\zeta^2\over \phi_-[\zeta\phi_-^2+(a+1)\phi_-
+\zeta]},\cr
F_3&(\zeta;-6\i,4,4(a-4)(a+2))=\i{2\zeta^2\phi_-^3+
(3a-2)\zeta\phi_-^2+ (2\zeta^2+a^2-a)\phi_- +(a-4)\zeta\over
\zeta\phi_-^2+(a-1)\phi_- +\zeta},\cr}$$
for $\epsilon_1=\epsilon_2=-\epsilon_3=-\epsilon_4=1$,
$\epsilon_1=\epsilon_2=-\epsilon_3=\epsilon_4=1$
and $\epsilon_1=\epsilon_2=\epsilon_3=-\epsilon_4=1$, respectively.

An analogous set of solutions may be obtained starting with the
one-parameter family solution of \piii\ expressed in terms of
modified Bessel functions, namely
$$\phi_\pm(\zeta;a-1,a+1,1,-1)={\d\over\d\zeta}\left\{
\ln\left[A\zeta^{-\nu}I_{\nu}(\pm \zeta)+B
\zeta^{-\nu}K_{\nu}(\pm \zeta)\right]\right\},\eqn{yMB}$$ where
$\nu=\tfr12(a-1)$,
$I_{\nu}(\zeta)$ and $K_{\nu}(\zeta)$ are Modified Bessel functions,
with $A$ and
$B$ are arbitrary constants. Applying (\yBTF) to this solution with
$\epsilon_3=\epsilon_4=-1$ yields
$$F(\zeta;-2\i,-4,-4(a-2)a)=\i[2\zeta\phi_\pm(\zeta) +a],$$
while applying it with $\epsilon_3=-\epsilon_4=-1$ yields
$$F(\zeta;-2\i
a,4,-4(a-2)(a+2))=\i{[2\zeta\phi_\pm^2(\zeta)+a\phi_\pm(\zeta)
-2\zeta]/\phi_\pm(\zeta)}\eqn{yMBsol}$$
Further solutions can be obtained by applying the \bt\ (\FBTscal).

All of the solutions of (\scalred) so far mentioned have been
imaginary.
In order to derive real solutions we make use of the
property that if $F(\zeta;\mu_1,\mu_2,K)$ is a solution of (\scalred)
then so is $F(\i\zeta;-\mu_1,-\mu_2,-K)$.
By letting $\i a=\alpha$, we obtain a new seed solution for the
\bt\ (\FBTscal) given
by $F_{r1}(\zeta;2\alpha,4,4(\alpha^2+1))=\pm2\zeta$,
and hence the following real rational solutions to (\scalred),
$$\eqalignno{
F_{r2}(\zeta;2\alpha,4,4(\alpha^2+9))=&\pm 2\zeta{4\zeta^2-4\alpha
\zeta+\alpha^2
-3\over 4\zeta^2 -4\alpha \zeta +\alpha^2+1},&\eqnm{Frat}{a}\cr
F_{r3}(\zeta;2\alpha,4,4(\alpha^2+25))=&\pm2\zeta {f_3(\zeta)\over
g_3(\zeta)},&\eqnr{b}\cr}$$ where
$$\eqalign{
f_3(\zeta)=64 \zeta^6&-192 \alpha \zeta^5+48 (5\alpha^2
-3)\zeta^4 -32\alpha (5 \alpha^2-7)\zeta^3 +60 (\alpha^4 -2
\alpha^2 -3)\zeta^2\cr &-12(\alpha^4-2 \alpha^2 -3)\alpha \zeta
+\alpha^6
-\alpha^4 +43\alpha^2 +45,\cr
g_3(\zeta)=64 \zeta^6 &-192 \alpha \zeta^5 +48(5 \alpha^2 +1) \zeta^4
-160(\alpha^2+1) \alpha \zeta^3 +12(5\alpha^4+14 \alpha^2+9)
\zeta^2 \cr &-12(\alpha^4+6\alpha^2 +5)\alpha \zeta +\alpha^6
+11\alpha^4+19\alpha^2 +9,\cr}$$
and so on. The rational solutions $F_{r2}(\zeta)$ and $F_{r3}(\zeta)$
are plotted in Figure 4. We note that Schief (1994) obtained the
special case
$\alpha=0$ of this hierarchy of rational solutions.

In a similar way the spherical Bessel function solution of order
$-\tfr12$ for
\piii\ given by
$$\phi(\zeta;-1,-1,1,-1)={[A\cos\zeta-B\sin\zeta]/[A\sin\zeta+B\cos\zeta]},$$
can be used to derive the hyperbolic solution of (\scalred),
$$F_{B1}(\zeta;0,4,16)=\pm{4\zeta\over\sinh^2 \zeta+\cosh^2
\zeta}=\pm{8\zeta\e^{2\zeta}\over\e^{4\zeta}+1},$$
where we have set $A=B$ in order for the solution to be real.
This solution can then be used in conjunction with \bt\ (\FBTscal)
to derive the following real solutions of (\scalred);
$$\eqalignno{
F_{B2}(\zeta;0,4,64)&=\pm{16\zeta\e^{2\zeta}\left[(2\zeta-1)\e^{4\zeta}
-2\zeta-1\right]\over
\e^{8\zeta}+2(8\zeta^2+1)\e^{4\zeta}+1},&\eqnm{Fbes}{a}\cr
F_{B3}(\zeta;0,4,144)&=\pm{8\zeta\e^{2\zeta}\left[
(8\zeta^2-12\zeta+3)\e^{8\zeta}-(64\zeta^4+32\zeta^2-6)
\e^{4\zeta}+8\zeta^2+12\zeta+3\right]\over
\e^{12\zeta}+(64\zeta^4-64\zeta^3+48\zeta^2+3)\e^{8\zeta}
+(64\zeta^4+64\zeta^3+48\zeta^2+3)\e^{4\zeta}
+1}.\cr&&\eqnr{b}\cr}$$ These solutions to (\scalred) are plotted
in Figure 5, where it can be seen that this family of solutions are
asymptotically zero as $\zeta \rightarrow \pm \infty$.

We can now use these solutions of (\scalred) to obtain hierarchies
of real solutions of (\eqgsg) by use of the formula, (\uvscal). Setting
$P(z,t)=\tfr12 \int^{zt^{1/2}} {F(s)}s^{-1}\d s$, we obtain from
(\Frat),
$$\eqalignno{P_{r1}(z,t)&=\pm \eta,&\eqnm{Prat}{a}\cr
P_{r2}(z,t)&=\pm \left\{\eta-2\tan^{-1} (2\eta-a)\right\},&\eqnr{b}\cr
P_{r3}(z,t)&=\pm\cases{\displaystyle\eta +
2\tan^{-1}\psi_1-2\tan^{-1}\psi_2-2\tan^{-1}\psi_3
+ 2 \tan^{-1}\left(2a\over a^2-1\right), & if $a\not=1$, \cr
\eta + 2\tan^{-1}\psi_4-2\tan^{-1}\psi_5-2\tan^{-1}\psi_3, & if $a=1$,
\cr}{\hbox to 20pt{\hfill}}&\eqnr{c}\cr}$$ where $\eta=zt^{1/2}$ and
$$\eqalign{
\psi_1&={8\over 3(a^{2}+1)}{\eta^5}-{8a\over
(a^{2}+1)}{\eta^4}+{4(a^{2}+7)\over 3(a^{2}+1)}{\eta^3}
-16a{\eta^2} +\tfr12(3a^{2}+7){\eta}-\tfr16(a^{2}+7)a,\cr
\psi_2&={\tfr{2}{3}}{\eta^3}-{\tfr{4}{3}a} \eta^2+
\tfr{5}{6}(a^2+1) \eta -{\tfr{1}{6}a^3}-{\tfr{7}{6}a},\cr
\psi_3&={\tfr{2}{3}}{\eta}-{\tfr{1}{3}a},\cr
\psi_4&={\tfr{4}{3}}{\eta^5}-4{\eta^4}+{\tfr{16}{3}}{\eta^3}-
{\tfr{16}{3}}{\eta^2}+5{\eta}-\tfr{4}{3},\cr
\psi_5&={\tfr{2}{3}}{\eta^3}-
{\tfr{4}{3}}{\eta^2}+{\tfr{5}{3}}{\eta}-\tfr{4}{3},\cr}$$

Similarly, we obtain from the family (\Fbes),
$$\eqalignno{
P_{B1}(z,t)&=\pm\tan^{-1}\left\{\sinh(2\eta)\right\},&\eqnm{Pbes}{a}\cr
P_{B2}(z,t)&=\pm\left\{\ln\left[\ex{4\eta}+4\i\eta\ex{2\eta}+1\right]
-\ln\left[\ex{4\eta}-4\i\eta\ex{2\eta}+1\right]\right\},
&\eqnr{b}\cr
P_{B3}(z,t)&=\pm\i\left\{\ln\left[\ex{6\eta}+\i(8\eta^2-4\eta+1)\ex{
4\eta}+(8\eta^2+4\eta+1)\ex{2\eta}+\i\right]\right.\cr
&\qquad\left.
-\ln\left[\ex{6\eta}-\i(8\eta^2-4\eta+1)\ex{4\eta}+(8\eta^2+4\eta
+1)\ex{2\eta}-\i\right]\right\},{\hbox to 20pt{\hfill}}&\eqnr{c}}$$
where
$\eta(z,t)=zt^{1/2}$.
We note again that the expressions (\Pbes) are real, despite their
closed
form expressions involving the complex number $\i$.
In Figure 6, we plot the solution
$u(x,y,t)=P_{B3}(x+y,t)+P_{B3}(x-y,t)$
for $t=1$, $4$ and $25$. In Figure 7, we plot the solution
$u(x,y,t)=P_{B3}(x+y,t)+P_{r3}(x-y,t)$ for $a=1$ and $t=1$.

\smallskip\noindent{\bf Case \secsub(ii)} $\mu_1\ne 0$, $\mu_2=0$.\quad
Given $\phi(\zeta)$ and $\~{\phi}(\tzeta)$ satisfying \piii\ (\eqPIII)
with
constants $\alpha_3=2\i{\mu_1 /\nu}$, $\beta_3=\tfr18{\nu}\left(4+{\i
K/\mu_1}\right)$, $\gamma_3=0$ and $\delta_3=-\tfr14\nu^2$, and
$\~\alpha_3=2\i{\~\mu_1/\~\nu}$, $\~\beta_3=\tfr18{\~\nu}\left(4+{\i
\~K/\~\mu_1}\right)$, $\~\gamma_3=0$ and $\~\delta_3=-\tfr14\~\nu^2$,
respectively, with $\nu\ne 0$ and $\~\nu\ne 0$, we have,
$$\eqalignno{
u(x,y,t)&=\tfr12{\i}\int^{\zeta}\left[{{\phi_{\zeta_1}(\zeta_1)}
-\tfr12\nu\over\phi(\zeta_1)}\right]\,\d\zeta_1
+\tfr12\i\int^{\tzeta}\left[{\~{\phi}_{\tzeta_1}(\tzeta_1)
-\tfr12\~\nu\over\~\phi(\tzeta_1)}\right]\,\d\tzeta_1,&\eqnm{equvIIIa}{a}\cr
v(x,y,t)&=-\tfr14\int^{\zeta}\int^{\zeta_1}\left[{{\phi_{\zeta_1}(\zeta_1)}
-\tfr12\nu\over\phi(\zeta_1)}\right]^2\d\zeta_2\,\d\zeta_1
-\tfr14\int^{\tzeta}\int^{\tzeta_1}\left[{\~{\phi}_{\tzeta_2}(\tzeta_2)
-\tfr12\~\nu\over\~\phi(\tzeta_2)}\right]^2\d\tzeta_2\,\d\tzeta_1,{\hbox
to
20pt{\hfill}} &\eqnr{b}\cr}$$
where $\zeta=(x+y)t^{1/2}$ and $\tzeta=(x-y)t^{1/2}$. The simplest
example of such a solution of \piii\ is given by $\phi(\zeta)=\epsilon
\zeta^{1/3}$
and $\~{\phi}(\tzeta)=\epsilon\tzeta^{1/3}$, in which case we have the
solution for
(\eqgsg), where $\cc1$ is determined by the parameters $\nu$,
$\epsilon$ and $\mu_1$,
$$\eqalign{
u(x,y,t)&=\tfr1{6}{\i}\ln(\zeta\tzeta)+
\cc1(\zeta^{2/3}+\tzeta^{2/3}),\cr
v(x,y,t)&=\tfr1{36}\ln(\zeta\tzeta)-\i\cc1\left(\zeta^{2/3}+\tzeta^{2/3}\right)
+\cc1^2\left(\zeta^{4/3}+\tzeta^{4/3}\right).}$$

A \bt\ for \piii\ with $\gamma\equiv0$ is given in equations (1.2a,b)
of
Milne \&\ Clarkson (1993). Using this \bt\ on the ``seed" solution
$\phi(\zeta)
=\epsilon \zeta^{1/3}$, one
can obtain a hierarchy of solutions of
\piii\ with $\gamma=0$ which are rational in $\zeta^{1/3}$.
We leave it to the reader to calculate the corresponding expressions
for
$u(x,y,t)$ and
$v(x,y,t)$, and the ``lift" of the \bt\ for \piii\ with $\gamma=0$
to the family of solutions (\equvIIIa) of (\eqgsg).

\smallskip\noindent{\bf Case \secsub(iii)} $\mu_1=\mu_2=0$. \quad
Given $\phi(\zeta;1,-\tfr14K,0,0)$ and
$\~{\phi}(\tzeta;1,-\tfr14\~K,0,0)$ satisfying
\piii\ (\eqPIII), we have,
$$\eqalignno{
u(x,y,t)&={\tfr12\i}\ln[\phi(\zeta)\~{\phi}(\tzeta)],&\eqnm{equvIIIb}{a}\cr
v(x,y,t)&=-\tfr14\int^{\zeta}\int^{\zeta_1}\left[{\phi'(\zeta_2)\over
\phi(\zeta_2)}
\right]^2\d\zeta_2\,\d\zeta_1 -
\tfr14\int^{\tzeta}\int^{\tzeta_1}\left[{\~{\phi}'(\tzeta_2)\over
\~{\phi}(\tzeta_2)}\right]^2\d\tzeta_2\,\d\tzeta_1
&\eqnr{b}\cr}$$
where $\zeta=(x+y)t^{1/2}$ and $\tzeta=(x-y)t^{1/2}$.
An alternative expression for $v$ is given by,
$$v(x,y,t)=-\tfr12\int^{\zeta}\left[\phi(\zeta_1) +{K\over
4\phi(\zeta_1)}\right]{\d\zeta_1\over \zeta_1} -\tfr12\int^{\tzeta}
\left[\~{\phi}(\tzeta_1) +{\~K\over
4\~{\phi}(\tzeta_1)}\right]{\d\zeta_1\over
\tzeta_1}.$$

In this case when $K=0$, the solution of \piii\ is
$$\eqalign{\phi(z)&={\cc1 z^{\lambda-1}\over\left(z^{\lambda}
-\tfr14\cc1\lambda^{-4}\right)^2},\qquad
\phi(z)={2\over z(\ln z+\cc1)^2}\cr}$$
(cf., Airault 1979). We leave it to the reader to calculate the
associated solutions
for (\eqgsg).

\section{Further solution families}
In this section we apply a known \bt\ for the system
(\eqgsg) to two of the simplest of the solutions found in the families
(\uvextra).
The result is new families of solutions involving the arguments and
moduli of sums
of products of exponential and Bessel function expansions with
arbitrary
coefficients, respectively. The analysis for the other exact solutions
described in
the previous section will be similar.

The \bt\ for the system (\eqgsg) we employ can be found in
Dubrovsky \&\ Konopelchenko (1993), Konopelchenko \etal\ (1992) and
Nimmo (1993).
Let $u$, $v$ be a solution of (\eqgsg). If $\psi$ is a solution of
$$\psi_{xy}-U\psi=0,\qquad U=u_xu_y-\i u_{xy}, \eqn{eqBTpsi}$$
then a new solution of (\eqgsg) is given by,
$$\~u=u+\i\ln(\psi/\bar\psi),\qquad\~v=v+\ln(\psi\bar\psi).\eqn{eqBT}$$
For the solution found in (\uvextra), we have
$$U=[\~\eta(\tzeta;\mu_1,1,\beta)]^2
+\~\eta_{\tzeta}(\tzeta;\mu_1,1,\beta)
-[\eta(\zeta;\mu_1,1,c)]^2 -{\eta_\zeta}(\zeta;\mu_1,1,c),$$  where
$\eta(\zeta;\mu_1,\mu_2,c)$ and
$\~\eta(\tzeta;\mu_1,\mu_2,\beta)$ satisfy (\wPII) with $\zeta=x+y-ct$
and
$\tzeta=x-y-\beta t$.  If we write,
$q(\zeta)=[\eta(\zeta;\mu_1,1,c)]^2+{\eta_\zeta}(\zeta;\mu_1,1,c)$ and
$\~q(\tzeta)=[\~\eta(\tzeta;\mu_1,1,\beta)]^2
+\~\eta_{\tzeta}(\tzeta;\mu_1,1,\beta)$, then we can rewrite (\eqBTpsi)
as,
$$\psi_{\zeta\zeta}-q(\zeta)\psi=\psi_{\tzeta\tzeta}
-\~q(\tzeta)\psi.$$
It is a curious fact that if
$f(r,s)=\psi_{rr}-q(r)\psi=\psi_{ss}+\~p(s)\psi$,
then $f(r,s)$ satisfies the same equation as does $\psi$.
Thus, if we can solve the eigenvalue problems for the ordinary
differential equations,
$$\phi_{\zeta\zeta}-q(\zeta)\phi=\lambda\phi,\qquad
\~\phi_{\tzeta\tzeta}-\~q(\tzeta)\~\phi=\lambda\~\phi,\eqn{eqlam}$$
we can set
$\psi(\zeta,\tzeta)=\phi(\zeta;\lambda)\,\~\phi(\tzeta;\lambda)$.
However we can go further; we can obtain solutions
of (\eqBTpsi) written as formal sums of eigenfunction expansions,
$$\psi(\zeta,\tzeta)=\sum_\lambda C_\lambda
\phi(\zeta;\lambda)\,\~\phi(\tzeta;\lambda),\eqn{eqpsisum}$$
where $C_\lambda$ is a constant and the $\psi_{i,\lambda}$ satisfy
(\eqlam) for
each given $\lambda$, since the governing equation for $\psi$,
(\eqBTpsi), is
linear.

As an example, we apply the \bt\
(\eqBT) to the ``seed'' solution, (\gsgsoli) which is obtained from the
solution
$Q(z;1) = {1/z}$ of (\wPII) with $\mu_2=1$, or from $Q=0$.
Then $p(s)=0=\~p(s)$, and the eigenvalue problem becomes
$$\phi_{i,zz}-\lambda\phi_i=0,\qquad\hbox{for}\qquad i=1,2.$$
If $\lambda=\mu^2$, then this has solution
$\phi_i(z) = A_\mu\exp(\mu\zeta)+B_\mu\exp(-\mu\zeta)$,
with $A_\mu$ and $B_\mu$ arbitrary constants, and so
$$\psi(\zeta,\tzeta)=\sum_\mu
\bigl[A_\mu\exp(\mu\zeta)+B_\mu\exp(-\mu\zeta)\bigr]
\bigl[\~A_\mu\exp(\mu\tzeta)+\~B_\mu\exp(-\mu\tzeta)\bigr].$$
If $\lambda=-\mu^2$, then
$\phi_i(z) =A_\mu\sin(\mu\zeta)+B_\mu\cos(\mu\zeta)$,
where $A_\mu$ and $B_\mu$ are constants, and so
$$\psi(\zeta,\tzeta)=\sum_\mu C_\mu\sin(\mu\zeta+\delta_\mu)
\sin(\mu\tzeta+\~\delta_\mu),$$
with $C_\mu$, $\delta_\mu$ and $\~\delta_\mu$ arbitrary constants.

Applying the \bt\ (\eqBT) to the solution (\gsgsolii) obtained from the
solution $\displaystyle y(x;2)={2(x^3-2)/[x(x^3+4)]}$ of \pii (\eqPII)
yields the
eigenvalue problem
$$\phi_{i,zz}+\left(\mu^2-2z^{-2}\right)\phi_i=0,\qquad i=1,2.$$
This has solution
$\phi_i(z) = z^{1/2}[A_\mu J_\nu(\mu z) + B_\mu J_{-\nu}(\mu z)]$,
where $A_\mu$ and $B_\mu$ are constants and $J_\nu(x)$ is the usual
Bessel function,
with $\nu=\tfr12\sqrt5$, and so
$$\psi(\zeta,\tzeta)=\sum_\mu \left(\zeta\tzeta\right)^{1/2}
\bigl[A_\mu J_\nu(\mu\zeta) + B_\mu J_{-\nu}(\mu\zeta)\bigr]
\bigl[\~A_\mu J_\nu(\mu\tzeta) + \~B_\mu J_{-\nu}(\mu\tzeta)\bigr].$$

It is clear that it is possible to obtain many further exact solutions
for (\eqgsg)
using solutions obtained in \S4 above in conjunction with the
\bt\ (\eqBT). We shall
not pursue this further here.

\section{Painlev\'e analysis and integrability of the generalized
Maxwell-Bloch
system} The {\it \p\ Conjecture\/} (or \p\ ODE test) as formulated by
Ablowitz \etal\
(1978, 1980), asserts that every \ode\ which arises as a symmetry
reduction of a
completely integrable nonlinear \pde\ is of \p\ type, though perhaps
only after a
trans\-forma\-tion of variables. Subsequently, Weiss \etal\ (1983),
proposed the
\p\ PDE test as a method of applying the \p\ ODE test directly to a
given \pde\
without having to consider symmetry reductions (which might not exist).
Despite
being by  no means foolproof the \p\ tests appear to provide a useful
criterion for
the identification of completely integrable \pdes.

\subsection{\p\ analysis of the Sine-Gordon system (\eqgsg)}
Here we discuss the integrability of the generalization of the
Sine-Gordon system
(\eqgsg) given by
$$\eqalignno{
&u_{xyt} + \alpha u_x v_{yt} + \beta u_y v_{xt}= 0,&\eqnm{eqggsg}{a}\cr
&v_{xy}- u_x u_y=0,&\eqnr{b}\cr}$$
where $\alpha$ and $\beta$ are arbitrary constants. To apply the
Painlev\'e PDE test
due to Weiss \etal\ (1983), to the system (\eqggsg) we seek a solution
in the form
$$u(x,y,t)=\sum_{k=0}^{\infty}u_k(y,t)\phi^{k+p}(x,y,t),\qquad
v(x,y,t)=\sum_{k=0}^{\infty}v_k(y,t)\phi^{k+q}(x,y,t),\eqn{eqVIi}$$
where $\phi(x,y,t)=x+\psi(y,t)$, with $\psi(y,t)$ an arbitrary analytic
function,
and $u_k(y,t)$, $v_k(y,t)$, $k=0,1,2,\ldots,$ are analytic functions
such that $u_0v_0\not\equiv0$, in the neighbourhood of an arbitrary,
non-characteristic movable singularity manifold defined by
$\phi(x,y,t)=0$,
and $p$ and $q$ are constants to be determined.
By leading order analysis we find that $p=q=0$. Thus the standard
procedure
for the applying the Painlev\'e PDE test to the system (\eqggsg) needs
to
be modified. Since  only the derivatives of $u$ and $v$ arise in
(\eqggsg), then,
following Clarkson (1986{\it b}), we seek a solution in the form
$$\eqalignno{u(x,y,t)&=u_{00}(y,t)\ln\phi(x,y,t)+\sum_{k=0}^{\infty}u_k(y,t)
\phi^{k}(x,y,t),&\eqnm{eqVIii}{a}\cr
v(x,y,t)&=v_{00}(y,t)\ln\phi(x,y,t)+\sum_{k=0}^{\infty}v_k(y,t)\phi^{k}(x,y,t),
&\eqnr{b}\cr}$$
where the same notation applies as in (\eqVIi) and $u_{00}(y,t)$ and
$v_{00}(y,t)$ are analytic functions to be determined.
By leading order analysis, from the coefficients of $\phi^{-3}$, we
find that
$u_{00}(y,t) = \i\sqrt{2/(\alpha+\beta)}$ and
$v_{00}(y,t)=2/(\alpha+\beta)$.
Equating the coefficients of powers of
$\phi^{r-3}$ yields the general recursion relation
$$\bss Q(r){u_r\choose v_r} \equiv
\pmatrix{r^2(r-3)\psi_y\psi_t & \sqrt{2(\alpha+\beta)}\,\i\, r
(r-1)\psi_y\psi_t\cr
\displaystyle{2\sqrt{2}\,\i\, r \psi_y/\sqrt{\alpha+\beta}} &
r(r-1)\psi_y\cr}{u_r\choose v_r} = {F_r\choose G_r},\eqn{}$$
where $F_r$ and $G_r$ are known functions of
$u_0,v_0,\ldots,u_{r-1},v_{r-1}$,
$\psi$ and their derivatives. These recursion relations uniquely define
$u_r$, $v_r$ unless $$\det \bss Q(r) =
(r+1)r^2(r-1)(r-4)\psi_y^2\psi_t=0.$$
Therefore the {\it resonances\/} are $r=-1,0,1,4$. The resonance $r=-1$
corresponds
to the fact that $\psi(y,t)$ is an arbitrary function and the double
resonance $r=0$
to both $u_0(y,t)$ and $v_0(y,t)$ being arbitrary functions.

For the resonance $r=1$, there is one equation whose solution yields
$u_1 =
-u_{0,y}/(2 \psi_y)$. Substituting this into the second equation yields
the
following ``compatibility condition'',
$$ (\alpha-\beta)\left(\sqrt{2}\,\psi_{yt} +
\i\sqrt{\alpha+\beta}\,u_{0,y}\psi_t\right)=0.$$
Since $\psi$ and $u_0$ are arbitrary, this is identically satisfied if
and only
if $\alpha=\beta$, in which case $v_1$ is another arbitrary function.
Otherwise, if $\alpha\not=\beta$, then it is necessary to introduce
$u_{11}(t)\phi(x,t)\ln\phi(x,t)$ and $v_{11}(t)\phi(x,t)\ln\phi(x,t)$
terms,
where $u_{11}(t)$ and $v_{11}(t)$ are to be determined, into the
expansion (\eqVIii)
and at higher orders of $\phi(x,t)$, higher and higher powers of
$\ln\phi(x,t)$
are required; a strong indication of non-\p\ behaviour.

If $\alpha=\beta$, then at the resonance $r=4$ there is only one
independent
equation and so either $u_4$ or $v_4$ is arbitrary. Hence we can indeed
obtain an
expansion for $u$, $v$ in the form (\eqVIii) with $u_0$, $v_0$, $v_1$
and $u_4$
(or $v_4$), arbitrary functions in the case when $\alpha=\beta$; since
the
expressions obtained are complex, we do not reproduce details here.
Therefore
the \p\ PDE test suggests that the Sine-Gordon system (\eqgsg) is the
only
integrable case of the system ({\eqggsg}).

\subsection{Integrability of the generalized, real, pumped
Maxwell-Bloch equation}
The integrability condition of the generalized, real, pumped
Maxwell-Bloch system (\eqws{\it a,b}) is
$$ww_{zzt}- w_zw_{zt}-m_1(t) w_{z}
+w^3w_{t}-\tfr12{\d m_2\over\d t}w^2=0,\eqn{eqwI}$$
where $m_1(t)$ and $m_2(t)$ are arbitrary functions.
It is routine to apply \p\ analysis to (\eqwI) and show that it
satisfies the
necessary conditions of the \p\ PDE test due to Weiss \etal\ (1983) to
be
completely integrable.

As mentioned in \S4.1 above, the sine-Gordon equation (\eqSG) is
equivalent to
the special case of (\eqwI) with $m_1(t)\equiv0$ and $m_2(t)\equiv0$,
whilst the
real, pumped Maxwell-Bloch system (\eqMBp) is equivalent to the special
case
with $m_1(t)\equiv0$. The ubiquitous sine-Gordon equation (\eqSG) is
one of the
fundamental soliton equations solvable by inverse scattering using the
AKNS
method (Ablowitz \etal\ 1974). However, the Lax pair for the complex
pumped
Maxwell-Bloch system is non-isospectral (Burtsev \etal\ 1987).

It can be shown that a non-isospectral Lax pair associated with (\eqwI)
is given by
$$\eqalignno{&\bfpsi_z = \left\{-\i\lambda\bfsigma_3 -
\tfr12\i w\bfsigma_2\right\}\bfpsi,&\eqnm{eqwlax}{a}\cr
&\bfpsi_t = \left\{\i{w_{zt}+m_1(t)\over
4\lambda w}\bfsigma_3  + {\i w_t\over 4\lambda}\bfsigma_1 +
{\i m_1(t)\over8\lambda^2}\bfsigma_2
\right\}\bfpsi =0.&\eqnr{b}\cr}$$
where $\bfsigma_1$, $\bfsigma_2$ and $\bfsigma_3$ are the Pauli spin
matrices
given by
$$\bfsigma_1=\pmatrix{\hfill 0 & \hfill 1 \cr 1 & 0 \cr},\qquad
\bfsigma_2=\pmatrix{0 & \hfill -\i \cr \i & \hfill 0 \cr},\qquad
\bfsigma_3=\pmatrix{1 & \hfill 0 \cr 0 & -1 \cr}.$$
Equations (\eqwlax{\it a,b}) are compatible, i.e.\
$\bfpsi_{zt}=\bfpsi_{tz}$ provided that $w(z,t)$ satisfies (\eqwI) and
$\lambda$
satisfies
$$\lambda^2(t)=-\tfr14m_2(t)+\lambda_0^2,\eqn{lambcond}$$ where
$\lambda_0$ is a
constant. Hence if $\displaystyle{\d m_2\over\d t}\not=0$, then the Lax
Pair (\eqwlax) is non-isospectral. We remark that the Lax pair
(\eqwlax) reduces to
that for the sine-Gordon equation (\eqSG) if
$m_1(t)=m_2(t)=0$ and that for the real, pumped Maxwell-Bloch system
(\eqMBp) if
$m_1(t)=0$. Further we note that the spectral problem (\eqwlax{\it a})
is the
standard AKNS spectral problem whilst if $m_1(t)\not=0$, then
(\eqwlax{\it b})
involves powers of both $\lambda^{-1}$ and $\lambda^{-2}$.

Thus, either there exists an isospectral Lax pair for (\eqwI), or the
definition of
integrability must be extended to include non-isospectral Lax pairs.
Either way, it
appears that the (\eqwI) can be viewed as an example of an equation on
the ``edge"
of integrability; it arises as a reduction of an integrable equation,
it possesses
the \p\ property yet only a non-isospectral Lax pair is currently
known.

\section{Discussion}
Using the nonclassical reduction method due to Bluman \&\ Cole (1969),
and a \bt\
(\eqBT), we are able to generate many families of exact solutions for
the system
(\eqgsg). Further, the reductions obtained have important consequences
for
the integrability of the pumped Maxwell-Bloch system, and the numerical
study of
(\eqgsg), as illustrated by Figure 3. The difficulty with using
numerics is in the
fact that every variable can be scaled arbitrarily and the result will
still be a
solution of the system; this is the content of the classical symmetry
group
for the system. Any numerical scheme for this equation must
be able to fix the scaling on each variable, otherwise a purely
artifactual
``chaos" will be observed.

There are two final points we wish to note. Firstly, the \bt\ (\eqBT)
for the
system (\eqgsg) appears to have no relation to the \bts\ discovered for
the various
families of exact solutions written down in \S4. Secondly, there now
exist several
examples where the nonclassical method has found a solution in the form
of a
decoupled sum of solutions of a lower dimensional equation
(cf.\ Clarkson \&\
Mansfield 1994{\it b}; Mansfield \&\ Clarkson 1994), a fact which is
interesting
in itself. Also of interest is the fact that there may exist families
of solutions
in which the pieces of the decoupled sum interact with each other.
Indeed, for the
shallow water wave equation discussed in Clarkson \&\ Mansfield
(1994{\it b}),
the nonclassical method obtained decoupled two-soliton solutions in
terms of
solutions of the first \p\ equation. The simplest examples of these
two-solitons
are part of a larger family of interacting two-solitons obtained by the
Hirota
method and by the singularity manifold method, but not by the
nonclassical
method. Conversely, neither the Hirota nor the singularity manifold
methods appear
to find the more general decoupled solution. It would be of great
interest if a
similar phenomenon were to be found for the system (\eqgsg).

\bigskip
{\baselineskip=12pt
\noindent {\bf Acknowledgement}. We thank Mark Ablowitz and Colin
Rogers, for
helpful comments and stimulating discussions. We also thank the Program
in
Applied Mathematics, University of Colorado at Boulder, for their
hospitality during
our visit whilst some of this work was done. The research of PAC and
ELM is
supported by EPSRC (grant GR/H39420) and that of AEM by an EPSRC
Research
Studentship, which are gratefully acknowledged.}

\def\refpp#1#2#3{{\rm#1}\ #2\ #3.}
\def\refjl#1#2#3#4#5#6{{\rm#1}\ #2\ #6\ {\frenchspacing\it#3}\ {\bf#4}
#5.}
\def\refbk#1#2#3#4{{\rm#1}\ #2\ ``{\sl#3},'' #4.}
\def\refcf#1#2#3#4#5#6#7{{\rm#1}\ #2\ #7\ in ``{\sl#3},''
({\rm\frenchspacing#4})\
#5, #6.}

\def\refn#1{\vskip 1pt\noindent\hangindent=20pt\hangafter=1}

\references
\baselineskip=12pt

\refn{refAC}
\refbk{Ablowitz, M.J.\ \& Clarkson, P.A.}{1991}{Solitons, Nonlinear
Evolution
Equations and Inverse Scattering}{{\frenchspacing\it L.M.S. Lect. Notes
Math.}, {\bf 149}, C.U.P., Cambridge}

\refn{refAbH}
\refjl{Ablowitz, M.J.\ \& Haberman, R.}{1975}{Phys. Rev.
Lett.}{35}{1185--1188}{Nonlinear evolution  equations --- two and three
dimensions}

\refn{refAKN}
\refjl{Ablowitz, M.J., Kaup, D.J.\ \& Newell,
A.C.}{1974}{\jmp}{15}{1852--1858}{Coherent pulse propagation, a
dispersive,
irreversible phenomenon}

\refn{refAKNS}
\refjl{Ablowitz, M.J., Kaup, D.J.,  Newell, A.C.\ \& Segur,
H.}{1974}{Stud.\
Appl.\ Math.}{53}{249--315}{The inverse scattering transform ---
Fourier
analysis for nonlinear problems}

\refn{refARSa}
\refjl{Ablowitz, M.J., Ramani, A.\ \& Segur,
H.}{1978}{\PRL}{23}{333--338}{Nonlinear  evolution equations and
ordinary
differential equations of \p\ type}

\refn{refARSb}
\refjl{Ablowitz, M.J., Ramani, A.\ \& Segur,
H.}{1980}{\jmp}{21}{715--721}{A
connection  between nonlinear evolution equations and ordinary
differential
equations of P-type. I}

 \refn{refAS}
\refbk{Ablowitz, M.J.\ \&\ Segur H.}{1981}{Solitons and the Inverse
Scattering Transform}{SIAM, Philadelphia}

\refn{refAir}
\refjl{Airault, H.}{1979}{\sam}{61}{31--53}{Rational solutions of
\p\ equations}

\refn{refAN}
\refjl{Athorne, C.\ \& Nimmo, J.J.C.}{1991}{Inverse
Problems}{7}{809--826}{On
the Moutard transformation for integrable partial differential
equations}

\refn{refBCa}
\refjl{Bluman, G.W.\ \& Cole, J.D.}{1969}{J.\ Math.\
Mech.}{18}{1025--1042}{The general similarity solution of the heat
equation}

\refn{refBLMP}
\refjl{Boiti, M., Leon, J.J.-P.,  Martina, L.\ \& Pempinelli,
F.}{1988}{Phys.
Lett.}{132A}{432--439}{Scattering of localized solitons in the plane}

\refn{refBur}
\refjl{Bureau, F.}{1972}{Ann. Mat. Pura Appl.
(IV)}{91}{163--281}{Equations
diff\'erentielles du second ordre en $\ddot y$ et du second degr\'e en
$y$
dont l'int\'egrale g\'en\'erale est \`a points critiques fix\`es}

\refn{refBurtsev}
\refjl{Burtsev, S.P.}{1993}{Phys.\ Lett.}{177A}{341--344}{The
Maxwell-Bloch
system with pumping and the fifth \p\ equation}

\refn{refBG}
\refjl{Burtsev, S.P.\ \& Gabitov, I.R.}{1994}{Phys.
Rev.}{A49}{2065--2070}{Alternative integrable equations of nonlinear
optics}

\refn{refBMZ}
\refjl{Burtsev, S.P., Mikhailov, A.V.\  \&  Zakharov,
V.E.}{1987}{Theor. Math.
Phys.}{70}{227--240}{Inverse scattering method with variable spectral
parameter}

\refn{refCHW}
\refjl{Champagne, B., Hereman, W.\ \& Winternitz, P.}{1991}{Comp.
Phys.
Comm.}{66}{319--340}{The computer calculation of Lie point symmetries
of large
systems of differential equations}

\refn{refPACi}
\refjl{Clarkson, P.A.}{1986{\it a}}{Lett. Math.
Phys.}{10}{297--299}{Remarks on
the two-di\-men\-sion\-al Sine-Gordon equation and the \p\ tests}

\refn{refPACii}
\refjl{Clarkson, P.A.}{1986{\it b}}{Physica}{19D}{447--450}{The \p\
property, a Modified Boussinesq equation and a Modified
Kadomtsev-Petviashvili
equation}

\refn{refPACiii}
\refjl{Clarkson, P.A.}{1990}{Europ. J. Appl. Math.}{1}{279--300}{New
exact
solutions of the Boussinesq equation}

\refn{refCMc}
\refjl{Clarkson, P.A.\ \& Mansfield,
E.L.}{1994{\it a}}{Physica}{70D}{250--288}{Symmetry reductions and
exact
solutions of a class of nonlinear heat equations}

\refn{refCMa}
\refjl{Clarkson, P.A.\ \& Mansfield, E.L.}{1994{\it
b}}{Nonlinearity}{7}{975--1000}{On a shallow water wave equation}

\refn{refCMb}
\refjl{Clarkson, P.A.\ \& Mansfield, E.L.}{1994{\it c}}{SIAM J.\ Appl.\
Math}{54}{to appear}{Algorithms for the nonclassical method of symmetry
reductions}

\refn{refCS}
\refjl{Cosgrove, C.M.\ \& Scoufis, G.}{1993}{\sam}{88}{25--87}{\p\
classification of a class of differential equations of the second order
and
second degree}

\refn{refDarboux}
\refpp{Darboux, G.}{1887--1896}{Le\c cons sur la th\'eorie g\'en\'erale
des surfaces et les applications g\'eom\'etrique du calcul
infinit\'esimal,
t.\ 1--4, Paris}

\refn{refKDb}
\refjl{Dubrovsky, V.G.\  \& Konopelchenko, B.G.}{1993}{Inverse
Problems}{9}{391--416}{The $2+1$-dimensional integrable generalization
of the
sine-Gordon equations: II. Localized solutions}

\refn{refFA}
\refjl{Fokas, A.S.\ \& Ablowitz, M.J.}{1982}{\jmp}{23}{2033--2042}{On a
unified
approach to transformations and elementary solutions of \p\ equations}

\refn{refFAb}
\refjl{Fokas, A.S.\  \& Ablowitz, M.J.}{1984}{\jmp}{25}{2494--2505}{On
the
inverse scattering transform of mul\-ti\-di\-men\-sion\-al nonlinear
evolution
equations related to first-order systems in the plane}

\refn{refFSi}
\refjl{Fokas, A.S.\ \& Santini, P.M.}{1989}{Phys. Rev.
Lett.}{63}{1329--1333}{Coherent structures in multidimensions}

\refn{refFSii}
\refjl{Fokas, A.S.\ \& Santini,
P.M.}{1990}{Physica}{44D}{99--130}{Dromions
and a boundary value problem for the Davey-Stewartson I equation}

\refn{refGNTZ}
\refjl{Gibbon, J.D., Newell, A.C., Tabor, M.\ \& Zeng, Y.B.}
{1988}{Nonlinearity}{1}{481--490}{Lax pairs, \bk\ transformations and
special
solutions for ordinary differential equations}

\refn{refGroma}
\refjl{Gromak, V.I.}{1975}{Diff. Eqns.}{11}{285--287}{Theory of \p's
equation}

\refn{refGromc}
\refjl{Gromak, V.I.}{1977}{Diff. Eqns.}{14}{1510--1513}{One-parameter
systems
of solutions of \p's equations}

\refn{refHere}
\refjl{Hereman, W.}{1994}{Euromath Bull.}{1{\rm ,
no.\ 2},}{45--79}{Review of
symbolic software for the computation of Lie symmetries of differential
equations}

\refn{refInce}
\refbk{Ince, E.L.}{1956}{Ordinary Differential Equations}{Dover, New
York}

\refn{refJRSS}
\refjl{Johnson, M.E., Rogers, C., Schief, W.K.\ \& Seiler,
W.M.}{1994}{Lie
Groups and their Applications}{1}{124--136}{On moving pseudospherical
surfaces:
a generalised Weingarten system and its formal analysis}

\refn{refKLM}
\refjl{Kitaev, A.V., Law, C.K.\  \& McLeod, J.B.}{1994}{Diff. Int.
Eqns.}{7}{967--1000}{Rational solutions of the fifth \p\ equation}

\refn{refKRT}
\refjl{Kitaev, A.V., Rybin, A.V.\  \& Timonen,
J.}{1993}{\jpa}{26}{3583--3595}{Similarity solutions of the deformed
Maxwell-Bloch system}

\refn{refKbk}
\refbk{Konopelchenko, B.G.}{1993}{Solitons in Multi-dimensions}{World
Scientific, Singapore}

\refn{refKDa}
\refjl{Konopelchenko, B.G.\ \& Dubrovsky, V.G.}{1993}{Stud. Appl.
Math.}
{90}{189--223}{The $2+1$-dimensional integrable generalization of the
sine-Gordon equations: I. $\partial$-$\=\partial$-dressing and the
initial value
problem}

\refn{refKR}
\refjl{Konopelchenko, B.G.\ \& Rogers, C.}{1991}{Phys.\
Lett.}{158A}{391--397}{On $2+1$-dimensional nonlinear systems of
Loewner type}

\refn{refKRb}
\refjl{Konopelchenko, B.G.\ \& Rogers, C.}{1993}{\jmp}{34}{214--242}{On
generalized Loewner systems: novel integrable equations in
$2+1$-dimensional}

\refn{refKSR}
\refjl{Konopelchenko, B.G., Schief, W.\ \& Rogers,
C.}{1992}{Phys.\ Lett.}{172A}
{39--48}{A $(2+1)$-dimensional sine-Gordon system: its auto-\bt}

\refn{refLW}
 \refjl{Levi, D.\  \& Winternitz,
 P.}{1989}{\jpa}{22}{2915--2924}{Nonclassical
symmetry reduction:  example of the Boussinesq equation}

\refn{refLoew}
 \refjl{Loewner, C.}{1952}{J. Anal. Math.}{2}{219--242}{Generation of
 solutions
of systems of partial differential equations by composition of
infinitesimal
\bts}

\refn{refLuka}
\refjl{Lukashevich, N.A.}{1965}{Diff.  Eqns.}{1}{561--564}{Elementary
solutions
of certain \p\  equations}

\refn{refMD}
\refpp{Mansfield, E.L.}{1993}{``{\tt diffgrob2}: A symbolic algebra
package for
analysing systems of PDE using {\sc maple}",
{\tt ftp euclid.exeter.ac.uk}, login: anonymous, password: your email
address,
directory: {\tt pub/liz}}

\refn{refMCa}
\refcf{Mansfield, E.L.\ \& Clarkson, P.A.}{1994}{Proceedings of the
Fourteenth
IMACS World Congress on Computation and Applied
Mathematics}{Ed.\ W.F.\ Ames}
{Georgia Inst.\ Tech.}{Vol.\ 1, pp336--339}{Applications of the
differential
algebra package {\tt diffgrob2} to reductions of PDE}

\refn{refMilne}
\refcf{Milne, A.E.\  \& Clarkson, P.A.}{1993}{Applications of Analytic
and
Geometric Methods to Nonlinear
Differential Equations}{Ed.\ P.A.\ Clarkson}{{\it NATO ASI Series C:
Mathematical and Physical Sciences\/}, {\bf 413}, Kluwer,
Dordrecht}{pp341--352}{Rational solutions and B\"acklund
transformations
for the third Painlev\'e equation}

\refn{refMurato}
\refjl{Murato, Y.}{1985}{Funkcial. Ekvacioj}{28}{1--32}{Rational
solutions of
the second and fourth \p\ equations}

\refn{refNimmp}
\refjl{Nimmo, J.J.C.}{1992}{Phys. Lett.}{168A}{113--119}{A class of
solutions of the Konopelchenko-Rogers equations}

\refn{refNimmo}
\refcf{Nimmo, J.J.C.}{1993}{Applications of Analytic and Geometric
Methods to
Nonlinear Differential Equations}{Ed.\ P.A.\ Clarkson}{{\it NATO ASI
Series C:
Mathematical and Physical Sciences\/}, {\bf 413}, Kluwer,
Dordrecht}{pp183--192}{Darboux transformations in $(2+1)$-dimensions}

\refn{refOkb}
\refjl{Okamoto, K.}{1987}{Funkcial. Ekvacioj}{30}{305--332}{Studies on
the \p\
equations. IV. Third \p\ equation ${\rm P}_{\rm III}$}

\refn{refOlver}
\refbk{Olver, P.J.}{1993}{Applications of Lie Groups to Differential
Equations}{Second Edition, Springer-Verlag, New York}

\refn{refRogers}
\refcf{Rogers, C.}{1993}{Applications of Analytic and Geometric Methods
to
Nonlinear Differential Equations}{Ed.\ P.A.\ Clarkson}{{\it NATO ASI
Series C:
Mathematical and Physical Sciences\/}, {\bf 413}, Kluwer,
Dordrecht}{pp217--228}{On gasdynamic-solitonic connections}

\refn{refSchiee}
\refjl{Schief, W.K.}{1992}{\jpa}{25}{L1351--L1354}{On localized
solitonic
solutions of a $(2+1)$-dimensional sine-Gordon system}

\refn{refSchief}
\refjl{Schief, W.K.}{1994}{\jpa}{27}{547--557}{\bts\ for the (un)pumped
Maxwell-Bloch system and the fifth \p\ equation}

\refn{refWTC}
\refjl{Weiss, J., Tabor, M.\  \& Carnevale,
G.}{1983}{\jmp}{24}{522--526}{The
\p\ property for partial differential equations}

\refn{refWW}
\refbk{Whittaker, E.E.\  \& Watson, G.M.}{1927}{Modern Analysis}{Fourth
Edition,
C.U.P., Cambridge}

\refn{refWint}
\refcf{Winternitz, P.}{1992}{\p\ Transcendents, their Asymptotics and
Physical
Applications}{Eds. P.\ Winternitz \&\ D.\ Levi}{{\it NATO ASI Series
B:
Physics\/}, {\bf 278}, Plenum, New York}{pp425--431}{Physical
applications of
\p\ type equations quadratic in the highest derivatives}

\figures
\noindent{\bf Figure 1.}
\hangindent=16pt\hangafter=1 The solution (\uvint{\it a}) where
$P=\tfr12\theta(x+y,2t)$, $\~{P}=\tfr12\theta(x-y,2\~{t})$
and $\theta$ is given by (\sgbre), with $\lambda=1$, $\mu=1.3$,
and $\~{t}=-2t$, for (i) $t=2$, (ii) $t=4$ and (iii) $t=6$.

\smallskip\noindent{\bf Figure 2.}
\hangindent=16pt\hangafter=1 The solution (\uvint{\it a}) where
$P=\tfr12\theta(x+y,2t)$, $\~{P}=\tfr12\theta(x-y,2\~{t})$,
and $\theta$ is given by (\sgkk) with $\beta_1=-1$, $\beta_2=2$, and
$\~{t}=2t$, for (i) $t=0$, (ii) $t=4$ and (iii) $t=8$.

\smallskip\noindent{\bf Figure 3.}
\hangindent=16pt\hangafter=1 The solution (\uvint{\it a}) where
$P=\tfr12\theta(x+y,2t)$, $\~{P}=\tfr12\theta(x-y,2\~{t})$ and $\theta$
is given by (\sgbre) with $\lambda=1$, $\mu=1.3$,
$y= 5$, but where $x$ has been scaled to $-\exp(1-x^2)$, $t$ has been
scaled to $\exp(1-t^2)$ and $\~t(t)=-\exp(1-t^2)$.

\smallskip\noindent{\bf Figure 4.}
\hangindent=16pt\hangafter=1 The solutions (\Frat{\it b,c}) to the
equation (\scalred) are plotted in (i) and (ii) respectively, with
$\alpha=1$.

\smallskip\noindent{\bf Figure 5.}
\hangindent=16pt\hangafter=1 The solutions (\Fbes{\it a,b,c})
to the equation (\scalred)
are plotted in (i), (ii) and (ii) respectively.

\smallskip\noindent{\bf Figure 6.}
\hangindent=16pt\hangafter=1 The solution (\uvscal{\it a}) where
$F$, $\^F$ are both given by (\Fbes{\it c}), shown for (i) $t=1$, (ii)
$t=4$ and (iii) $t=25$.

\smallskip\noindent{\bf Figure 7.}
\hangindent=16pt\hangafter=1 The solution (\uvscal{\it a}) where
$F$ is given by (\Fbes{c}) and $\^F$ is given by (\Frat{\it c}),
at $t=1$.
\vfill\eject\bye
{\vbox to 4.2 truein{\vfill}}
\centerline{\bf Figure 1(i)}
\vfill
\centerline{\bf Figure 1(ii)}
\eject

{\vbox to 4.2 truein{\vfill}}
\centerline{\bf Figure 1(iii)}
\vfill
\centerline{\bf Figure 2(i)}
\eject

{\vbox to 4.2 truein{\vfill}}
\centerline{\bf Figure 2(ii)}
\vfill
\centerline{\bf Figure 2(iii)}
\eject

{\vbox to 4.2 truein{\vfill}}
\centerline{\bf Figure 3}
\vfill
\centerline{\bf Figure 4(i)}
\eject

{\vbox to 4.2 truein{\vfill}}
\centerline{\bf Figure 4(ii)}
\vfill
\centerline{\bf Figure 5(i)}
\eject
{\vbox to 4.2 truein{\vfill}}
\centerline{\bf Figure 5(ii)}
\vfill
\centerline{\bf Figure 5(iii)}
\eject
{\vbox to 4.2 truein{\vfill}}
\centerline{\bf Figure 6(i)}
\vfill
\centerline{\bf Figure 6(ii)}
\eject
{\vbox to 4.2 truein{\vfill}}
\centerline{\bf Figure 6(iii)}
\vfill
\centerline{\bf Figure 7}
\eject
\end